\documentclass[11pt]{article}
\usepackage{epsfig} 
\usepackage{amssymb}
\newcommand{\sect}[1]{ \section{#1} \setcounter{equation}{0} }

\newcommand{\half}{\mbox{\small{$\frac{1}{2}$}}} 
\newcommand{\third}{\mbox{\small{$\frac{1}{3}$}}}

\newcommand{\pitwo}{\mbox{\small{$\frac{\pi}{2}$}}} 
\newcommand{\pisix}{\mbox{\small{$\frac{\pi}{6}$}}} 
\newcommand{\MSbar}{\overline{\mbox{MS}}} 
\newcommand{\MSbars}{\overline{\mbox{\footnotesize{MS}}}} 
\newcommand{\MOMg}{\mbox{MOMggg}}
\newcommand{\MOMgs}{\mbox{\footnotesize{MOMggg}}}
\newcommand{\MOMh}{\mbox{MOMh}}
\newcommand{\MOMhs}{\mbox{\footnotesize{MOMh}}}
\newcommand{\MOMq}{\mbox{MOMq}}
\newcommand{\MOMqs}{\mbox{\footnotesize{MOMq}}}
\newcommand{\mMOM}{\mbox{mMOM}}

\newcommand{\MOMi}{\mbox{MOMi}}
\newcommand{\MOMis}{\mbox{\footnotesize{MOMi}}}
\newcommand{\Nc}{N_{\!c}}
\newcommand{\Nf}{N_{\!f}}
\newcommand{\NF}{N_{\!F}}
\newcommand{\NA}{N_{\!A}}

\begin{document}
\title{Banks-Zaks fixed point analysis in momentum subtraction schemes}
\author{J.A. Gracey \& R.M. Simms, \\ Theoretical Physics Division, \\ 
Department of Mathematical Sciences, \\ University of Liverpool, \\ P.O. Box 
147, \\ Liverpool, \\ L69 3BX, \\ United Kingdom.} 
\date{}
\maketitle 

\vspace{5cm} 
\noindent 
{\bf Abstract.} We analyse the critical exponents relating to the quark mass
anomalous dimension and $\beta$-function at the Banks-Zaks fixed point in
Quantum Chromodynamics (QCD) in a variety of representations for the quark in 
the momentum subtraction (MOM) schemes of Celmaster and Gonsalves. For a 
specific range of values of the number of quark flavours, estimates of the 
exponents appear to be scheme independent. Using the recent five loop modified 
minimal subtraction ($\MSbar$) scheme quark mass anomalous dimension and 
estimates of the fixed point location we estimate the associated exponent as 
$0.263$-$0.268$ for the $SU(3)$ colour group and $12$ flavours when the quarks 
are in the fundamental representation.  

\vspace{-17cm}
\hspace{13.5cm}
{\bf LTH 1041}

\newpage 

\sect{Introduction.}

Non-abelian gauge theories are known to be asymptotically free field theories
due to the obseravtions made in \cite{1,2}, that for a certain range of the 
number of quark flavours, $\Nf$, the one loop $\beta$-function is negative. 
The main theory where this fundamental property is relevant is Quantum 
Chromodynamics (QCD) which is believed to underpin our understanding of 
nature's strong nuclear force. Certainly at high energy where the quarks and 
gluons of QCD behave as effectively free fundamental particles, this asymptotic
freedom feature has meant that the internal structure of hadrons can be probed 
experimentally. The range of $\Nf$ where asymptotic freedom is valid is limited
since when $\Nf$ is sufficiently large the one loop $\beta$-function becomes 
positive and one in effect is in a theory with properties similar to Quantum 
Electrodynamics (QED). One immediate question which arose in light of the one 
loop QCD $\beta$-function's emergence is whether the only perturbative fixed 
point of the $\beta$-function was the one at the origin. Insight into this 
problem was given after the computation of the two loop correction to 
$\beta(a)$, \cite{3,4}, where $a$~$=$~$g^2/(16\pi^2)$ is the perturbative 
coupling constant expressed in terms of the coupling constant $g$ of the quark 
and gluon interaction in the QCD Lagrangian. As initially discussed in \cite{3}
and further developed in detail in \cite{5}, the $\beta$-function can exhibit a 
non-trivial zero for a range of $\Nf$. This occurs when the one loop 
$\beta$-function is negative but the two loop coefficient is positive. Known 
now as the Banks-Zaks fixed point it has been studied since its discovery due 
to its potential connection with chiral symmetry breaking, for instance. In 
more recent years interest in this fixed point has in the main been due to the 
connection with physics beyond the Standard Model such as technicolor,
\cite{6,7}. More specifically while the early focus was on QCD itself, taking 
colour groups other than $SU(3)$ with quarks in non-fundamental representations
opened up the analysis to model building. This is primarily due to the need to 
understand where the conformal window is and the {\em true} range for which it 
exists. By conformal window we mean the range of $\Nf$ for which the 
non-trivial fixed point exists. The need to find the true range is not a 
trivial statement. The original observation of \cite{5} used the two loop 
$\beta$-function and this implicitly assumed that the Banks-Zaks fixed point 
was accessible perturbatively which is not necessarily the case. The problem is
that at the lower end of the conformal window, which for QCD is $\Nf$~$=$~$9$,
\cite{3}, the location of the fixed point is beyond the range of perturbative 
reliability. So while there may be a formal non-trivial zero of $\beta(a)$, 
there is no rigorous evidence that it truly exists for relatively low values of
$\Nf$. Only a non-perturbative analysis could resolve this. In this respect 
there has been intense interest in the lattice community in studying this 
problem for relatively large values of $\Nf$ but which are on the limit of 
perturbative reliability. A non-exhaustive representation of such lattice 
analyses can be found in \cite{8,9,10,11,12,13,14,15}, for example. Though 
studies have also been performed with Schwinger-Dyson methods, \cite{16}. More 
specifically $\Nf$~$=$~$12$ QCD lattice measurements have been made 
\cite{9,14,15}. Part of the motivation is to understand how to find the 
non-trivial fixed points non-perturbatively and from that knowledge endeavour 
to explore the fixed point structure for values of $\Nf$~$\leq$~$6$, if it 
exists, in order to tackle the relation to chiral symmetry breaking. 

One of the main topics of current analyses is the measurement of critical
exponents associated with the phase transition corresponding to the Banks-Zaks 
fixed point. These can be determined relatively accurately on the lattice. 
Indeed several recent studies, \cite{9,14,15}, show good agreement for the 
$\Nf$~$=$~$12$ quark mass anomalous dimension exponent. This exponent is of 
primary interest because of its relation to the definition of a conformal 
theory. Briefly the full dimension of the quark mass operator must be larger
than unity. This places an upper bound of $2$ on the contribution of the 
{\em anomalous} dimension to this for the theory to be conformal. (See, for
example, the discussion in \cite{17}.) Determining the range of the conformal 
window for which a theory satisfies this condition is an indication of the 
whether conformal symmetry is present. However, the determination of critical 
exponents is not limited to the lattice. They can be computed from knowledge of
the renormalization group functions. As in \cite{5} the explicit location of 
the fixed point can be deduced numerically and then the renormalization group 
functions are evaluated at that point to give estimates for the exponents. In 
the intervening years after the two loop work of \cite{3,4}, the $\MSbar$ QCD 
$\beta$-function has been extended to four loops as has the quark mass 
anomalous dimension, \cite{18,19,20,21,22,23,24,25,26,27}. With this higher 
order information the location of the Banks-Zaks fixed point has been refined. 
See, for example, \cite{28}. At a more technical level the work of 
\cite{29,30,31,32,33} formally examines the dependence of the fixed point 
structure in various schemes and finds conditions on the relations between 
schemes which ensure credible results. More recently a comprehensive explicit 
study has been provided in \cite{17}. There a range of colour groups has been 
examined with quarks in various representations which are relevant to several 
problems such as those underlying technicolor theories. One general feature of 
the results of \cite{17} was that the exponent estimates were becoming more 
reliable when higher order perturbation theory was taken into account. Indeed 
there was an indication that a selection of estimates were converging. Although
whether this was to a value which would be competitive with lattice estimates 
was not entirely clear for values of $\Nf$ in the low to mid-range of the 
conformal window. It would not be surprising if they did not 
since non-perturbative properties are present within lattice regularized 
theories. In more detail the perturbative analysis of \cite{17} provided 
estimates for $\Nf$~$=$~$12$ on the edge of the error ranges given on the 
lattice, \cite{14,15}. One important guide to the credibility of exponent 
estimates using the renormalization group function approach was the analysis in
schemes other than $\MSbar$, \cite{17,34,35}. It is a property of the critical 
point renormalization group equation that critical exponents are 
renormalization group invariants. Therefore, the value one obtains for an 
exponent is independent of the renormalization scheme used to perform the 
computations. Of course, this is in the ideal scenario where one knows the 
renormalization group functions to all orders in various schemes. This is not 
the situation in general. So by computing in various schemes for {\em four}
dimensional theories, as was carried out in \cite{17}, it may be the case that 
the convergence is faster than compared to another scheme. Although one never 
knows a priori which if any scheme would have this property. In \cite{17} the 
schemes which were considered were the $\MSbar$, modified regularization 
invariant (RI${}^\prime$), \cite{36,37}, and minimal momentum subtraction 
($\mMOM$) schemes, \cite{38}. The renormalization group functions for the final
two schemes are also known at four loops, \cite{36,37,38,39,40,41}. However, in 
some sense the three schemes are similar being defined with respect to Green's 
functions where there is a nullified external momentum. The $\mMOM$ scheme, for
example, is based on the property that in the Landau gauge the ghost-gluon 
vertex is finite, \cite{42}, when one ghost external leg is nullified. This 
feature allows one to assign a scheme for an {\em arbitrary} linear covariant 
gauge. For a $3$-point function this nullified external momentum configuration 
is termed an exceptional configuration and hence has potential infrared 
issues. While this ought not to be a problem for high energy analysis one has 
to be cautious in any low energy studies. 

In \cite{43,44} an alternative set of renormalization schemes was introduced
where the $3$-point QCD vertices were renormalized at a non-exceptional
external momentum configuration known as the symmetric point. Three momentum 
subtraction schemes (MOM) were defined based on the triple gluon, ghost-gluon 
and quark-gluon vertices and denoted respectively by $\MOMg$, $\MOMh$ and 
$\MOMq$, \cite{43,44}. By their very nature they are 
physical schemes which are mass dependent. In \cite{44} one hope which was
expressed was that perturbative results in the MOM schemes would have a faster 
convergence than other schemes. What is perhaps more relevant, however, is that
there is no doubt about infrared issues due to the non-exceptionality of the 
subtraction point. In light of this and interest in the Banks-Zaks fixed point 
the aim of this article is to extend the analysis of \cite{17} to the three MOM
schemes of \cite{43,44}. This is possible partly due to the provision of the 
three loop MOM $\beta$-functions, \cite{45}, for an arbitrary linear covariant 
gauge. Although our main interest here will be in the Landau gauge. At this 
loop order the scheme dependence first appears in this latter gauge and so it 
is apt to study the convergence and scheme dependence of the Banks-Zaks 
critical exponents in another set of schemes. As will be evident from the 
explicit structure of the expressions for the renormalization group functions 
the MOM schemes are in a different class to those used in \cite{17}. So these 
MOM schemes will offer non-trivial insight into properties of the Banks-Zaks 
fixed point discussed here. In order to carry out our study the first 
ingredient is to determine the quark mass anomalous dimension in the MOM 
schemes at three loops. These were not constructed in \cite{45} and require the
renormalization of the quark mass operator inserted in a two loop quark 
$2$-point function where there is a non-zero external momentum flow through the
inserted operator. 

The article is organized as follows. We derive the three loop quark mass
anomalous dimension in the three MOM schemes in section $2$ by exploiting
properties of the renormalization group equation. Properties of fixed points
are reviewed in section $3$ such as the renormalization group invariance of
critical exponents. In particular we show that the critical exponents derived
in the $\MSbar$ and MOM schemes at the Wilson-Fisher fixed point are the
same to the various loop orders to which they are known. This clarifies why
renormalization group functions, which in different schemes have different 
analytic structure, do produce renormalization group invariant Wilson-Fisher 
fixed point exponents. Our extensive Banks-Zaks analysis is provided in section
$4$. In the main the results are collected across various tables for ease of
viewing. Conclusions are given in section $5$.

\sect{Mass operator anomalous dimension.}

We begin our analysis by determining the quark mass anomalous dimension in the
three MOM schemes using the same approach of others in a chiral theory, 
\cite{25,46}. Rather than renormalize the mass itself directly its anomalous 
dimension is deduced from the renormalization of the associated quark mass 
operator which is $\bar{\psi} \psi$. This is renormalized by inserting it into 
a quark $2$-point function and ensuring that that Green's function is rendered 
finite with respect to the particular renormalization scheme of interest. For 
instance, in \cite{25} the original three loop $\MSbar$ renormalization 
constant for the quark mass operator was inserted at zero momentum in this 
Green's function. This was the appropriate external momentum configuration for 
this particular scheme since one is only interested in the divergences with 
respect to the regularizing parameter. Throughout we will use dimensional 
regularization in $d$~$=$~$4$~$-$~$2\epsilon$ dimensions with $\epsilon$ being 
the regularizing parameter. The advantage of this momentum configuration in the
derivation of the results of \cite{25,46} is that the Green's function in 
effect reduces to the computation of massless $2$-point Feynman diagrams which 
are readily calculable by standard techniques such as the {\sc Mincer} 
algorithm, \cite{47}. Although the original three loop results used the $R$ 
operation and infrared rearrangement of \cite{48,49}, later three loop 
computations of quark bilinear operator anomalous dimensions used {\sc Mincer},
\cite{46}. The subsequent extensions of the three loop result, \cite{26,27,50},
have used several different approaches. In \cite{26} an adaptation of 
{\sc Mincer} was developed which used a posteriori the four loop massless 
master $2$-point functions of \cite{51} while infrared rearrangement, 
\cite{48,49}, together with the evaluation of four loop massive vacuum bubble 
graphs was used in \cite{27}. While such methods of reducing the 
renormalization of the quark mass operator to $2$-point functions allows access
to higher order $\MSbar$ anomalous dimensions the particular external momentum 
configuration which was used, which is exceptional, cannot be exploited for the
set of MOM schemes of \cite{43,44}. They require a momentum configuration where
there is a non-zero momentum flowing through all external legs which means the 
configuration is non-exceptional. Hence it should suffer none of the infrared 
problems that could potentially arise in the {\sc Mincer} approach. Though we 
need to qualify these remarks briefly. First, the computations of \cite{25,46} 
are perfectly infrared safe through the use of infrared rearrangement, 
\cite{48,49}. Also the {\sc Mincer} package has actually been used to study 
symmetric point vertex functions in \cite{52}. However, this used {\sc Mincer} 
to approximate the basic integrals numerically rather than analytically by an
expansion method. Nevertheless compared to the exact three loop MOM 
$\beta$-functions which were determined in \cite{45} there was agreement to a 
few percent. Therefore, to determine the quark mass operator in the MOM schemes
of \cite{43,44} we have to consider the Green's function
\begin{equation}
\left. \left\langle \psi(p) \bar{\psi}(q) [\bar{\psi} \psi] (r)
\right\rangle \right|_{p^2 = q^2 = - \mu^2} 
\label{greenfn}
\end{equation}
where
\begin{equation}
p ~+~ q ~+~ r ~=~ 0
\end{equation}
and $p$, $q$ and $r$ are the three external momenta. We will always take $p$
and $q$ as the two independent momenta. The restriction in (\ref{greenfn})
indicates evaluation at the symmetric point which is defined as
\begin{equation}
p^2 ~=~ q^2 ~=~ r^2 ~=~ -~ \mu^2
\end{equation}
implying
\begin{equation}
pq ~=~ \frac{1}{2} \mu^2 ~.
\end{equation}
Here $\mu$ is the mass scale which is introduced to ensure that with 
dimensional regularization the coupling constant, denoted by $g$ here, is
dimensionless in $d$-dimensions. In keeping with previous work we retain the
same conventions here which were used in \cite{45}. 

The evaluation of (\ref{greenfn}) requires some care since we will be using the
same computational algorithm as \cite{45} to determine the Green's function.
First, (\ref{greenfn}) has to be decomposed into its Lorentz scalar components.
For the symmetric point there are two possible independent Lorentz tensors in
this basis which are 
\begin{equation}
{\cal P}^{\bar{\psi} \psi}_{(1)}(p,q) ~=~ \Gamma_{(0)} ~~,~~
{\cal P}^{\bar{\psi} \psi}_{(2)}(p,q) ~=~ \Gamma_{(2)}^{pq}
\end{equation} 
where
\begin{equation}
\Gamma_{(n)}^{\mu_1 \ldots \mu_n} ~=~ \gamma^{[\mu_1} \ldots \gamma^{\mu_n]}
\end{equation}
are totally antisymmetric generalized $\gamma$-matrices discussed in 
\cite{53,54,55}. The normalization of $1/n!$ is included in the definition. 
This specific choice of $\gamma$-matrices means that the spinor space into 
which (\ref{greenfn}) decomposes partitions due to, \cite{53,54,55},
\begin{equation}
\mbox{tr} \left( \Gamma_{(m)}^{\mu_1 \ldots \mu_m}
\Gamma_{(n)}^{\nu_1 \ldots \nu_n} \right) ~ \propto ~ \delta_{mn}
I^{\mu_1 \ldots \mu_m \nu_1 \ldots \nu_n}
\end{equation}
where the unit matrix is denoted by 
$I^{\mu_1 \ldots \mu_m \nu_1 \ldots \nu_n}$. We use the convention that when a
Lorentz index is contracted with a momentum then the dummy index is replaced by
that momentum. Clearly for the momentum configuration which was used to derive
the original $\MSbar$ high loop quark mass anomalous dimension one would have 
only one tensor in its decomposition basis since then $p$ and $q$ would be 
parallel. Therefore, for the symmetric point evaluation we define the 
projection by 
\begin{equation}
\left. \left\langle \psi(p) \bar{\psi}(q) [\bar{\psi} \psi] (r)
\right\rangle \right|_{p^2 = q^2 = - \mu^2} ~=~ 
\sum_{k=1}^{2}
{\cal P}^{\bar{\psi} \psi}_{(k)}(p,q) \,
\Sigma^{\bar{\psi} \psi}_{(k)}(p,q)
\end{equation}
where $\Sigma^{\bar{\psi} \psi}_{(k)}(p,q)$ are values of the scalar amplitudes
at the symmetric point. To determine these explicitly we use the projection 
method of \cite{45} where formally
\begin{equation}
\Sigma^{\bar{\psi} \psi}_{(k)}(p,q) ~=~
{\cal M}^{\bar{\psi} \psi}_{kl} \left. \left(
{\cal P}^{\bar{\psi} \psi}_{(l)}(p,q) 
\left\langle \psi(p) \bar{\psi}(q) [\bar{\psi} \psi](r)
\right\rangle \right) \right|_{p^2 = q^2 = - \mu^2}
\end{equation}
and
\begin{equation}
{\cal M}^{\bar{\psi} \psi} ~=~ \frac{1}{12} \left(
\begin{array}{cc}
3 & 0 \\
0 & - 4 \\
\end{array}
\right) ~.
\end{equation}
Applying this projection to each of the Feynman graphs comprising the Green's
function produces scalar integrals involving scalar products of the external
and internal momenta. 

To evaluate these we used the Laporta approach, \cite{56}, where an intense 
amount of integration by parts produces a small set of basic master integrals. 
These have been computed explicitly over several years in \cite{57,58,59,60}
but we use the notation of \cite{61} where there is a summary of the master 
values in powers of $\epsilon$ to the order required to determine the finite 
part. In practical terms we used the version of the Laporta algorithm which was 
implemented in the {\sc Reduze} package, \cite{62}. One useful feature of that
package is that it creates a large database of relations between the integrals 
and solves them automatically in terms of the masters. The relations and 
results necessary for the computation at hand can be readily lifted from the 
database and converted into {\sc Form} notation. We use {\sc Form} and its 
threaded version {\sc Tform}, \cite{63,64}, as the medium to handle the tedious 
amounts of large algebra which arise in the evaluation of the Green's function.
Indeed this was the approach used in similar previous work, \cite{45}. The 
Feynman diagrams contributing to (\ref{greenfn}) are generated with 
{\sc Qgraf}, \cite{65}. At one loop there is only one graph while at two loops 
there are $13$ graphs. Once all the necessary components of this algorithm are 
assembled the calculation runs automatically. Included in this is the way we 
undertake the renormalization which follows the method of \cite{20}. The 
Green's function is determined as a function of the bare coupling constant and 
gauge parameter but their respective counterterms are introduced by replacing 
the bare quantities by the renormalized parameters. The renormalization 
constant associated with each produces the canonical counterterms at each order
in perturbation theory. The remaining overall divergences, as well as the 
appropriate finite part in the MOM case, are finally absorbed into the overall 
renormalization constant for the Green's function. In this case this will be 
the quark mass operator renormalization constant. 

As part of this renormalization discussion it is worth defining the MOM schemes
for the quark mass anomalous dimension, $\gamma_{\bar{\psi}\psi}(a,\alpha)$
where $\alpha$ is the gauge parameter of the canonical linear covariant gauge
fixing. First, to carry out an $\MSbar$ determination of 
$\gamma_{\bar{\psi}\psi}(a,\alpha)$ for the symmetric point momentum 
configuration only the poles of the Green's function are important. However, 
the wave function renormalization of the external quark fields has to be 
included which will be the two loop $\MSbar$ ones of \cite{18,66}. Following 
this procedure we have verified that the two loop $\MSbar$ value of  
$\gamma_{\bar{\psi}\psi}(a,\alpha)$ is obtained. This is a check on our 
computer algebra set-up as the original two loop computation of \cite{24},
as well as that of \cite{23}, was performed by the direct evaluation of the 
quark $2$-point function in the presence of {\em massive} quarks. Having 
verified this for (\ref{greenfn}) then we can repeat the computation for the 
various MOM schemes. This is similar in each case but requires not only the 
quark wave function renormalization constant but also the gauge parameter and 
coupling constant renormalization constants all in the same MOM scheme. The 
explicit values in each of the three schemes for these quantities are given in
\cite{43,44,45}. We note that in \cite{45} the gauge parameter renormalization 
is performed in a MOM way. In some symmetric point analyses this parameter is 
renormalized in an $\MSbar$ fashion. However, as we are ultimately only 
interested in the expressions in the Landau gauge then the differences between 
the anomalous dimensions in both approaches would only be apparent in the 
$\alpha$ dependent terms. In other words they would be equivalent in the Landau
gauge. Any expression we present here which depends explicitly on $\alpha$ will
have used a MOM definition for the renormalization of $\alpha$. The main reason
we retain it within our computations is mainly as an internal check. For 
example, in the $\MSbar$ scheme the quark mass anomalous dimension is 
independent of $\alpha$ as the operator is gauge invariant. So we have checked 
that the two $\alpha$ independent mass operator renormalization constant 
correctly emerges when we compute in an arbitrary linear covariant gauge. We 
note that the full analytic expressions for all main results here are provided 
in an attached electronic data file.

The main reason why we concentrate on the Landau gauge is due to the 
renormalization group. The gauge parameter of the linear covariant gauge fixing
appears in the $\MSbar$ and MOM renormalization group functions and can be
regarded as a second coupling constant albeit to a quadratic term in the gauge
field. In this set of gauges the gauge parameter anomalous dimension can be
thought of as the $\beta$-function of $\alpha$. Thence it has in principle to
be included in any fixed point analysis. Clearly from the high order loop
anomalous dimension for $\alpha$ in the various schemes the anomalous dimension
is proportional at $\alpha$. So that $\alpha$~$=$~$0$ is a fixed point and
hence the focus on the Landau gauge. Of course, this is not the only solution
since in principle there could be non-trivial Banks-Zaks type fixed points for 
$\alpha$ itself. We do not consider those here partly because the lattice
analyses are in the Landau gauge. Some insight, though, into such additional
fixed points has been given in \cite{29,30,31,32,67}.

While what we have described is the procedure to construct the two loop MOM
operator renormalization constants, the {\em three} loop anomalous dimension in
each of the MOM schemes can be determined with this information. This is 
possible due to a property of the renormalization group equation and knowledge
of the three loop $\MSbar$ quark mass anomalous dimension, \cite{25}. The
construction requires the operator renormalization conversion function which is
defined by 
\begin{equation}
C^{\MOMis}_{\bar{\psi}\psi} (a,\alpha) ~=~ Z^{\bar{\psi}\psi}_{\MOMis}
\left[ Z^{\bar{\psi}\psi}_{\MSbars} \right]^{-1} ~.
\label{masscon}
\end{equation}
In (\ref{masscon}) the convention we use is that the function is expressed in
terms of $\MSbar$ variables for the coupling constant and gauge parameter. We
do not include the $\MSbar$ label on these variables. However, in computing the
right hand side of (\ref{masscon}) each renormalization constant is a function
of the parameters defined in those respective schemes. In order to have a
finite function in the $\epsilon$~$\rightarrow$~$0$ limit the MOM variables
have to be mapped to their $\MSbar$ versions before the perturbative
expansion of $C^{\MOMis}_{\bar{\psi}\psi} (a,\alpha)$ is deduced. For each of
the MOM schemes we are interested in here these mappings are given in
\cite{45}. The full expressions for the quark mass conversion function is given
in the associated data file. However, the numerical expression in each MOM 
scheme for $SU(3)$ is 
\begin{eqnarray}
C^{\MOMis}_{\bar{\psi}\psi}(a,\alpha) &=& 
1 ~+~ [ 0.229271 \alpha - 0.645519 ] a \nonumber \\
&& +~ [ 0.568426 \alpha^2 + 4.554664 \alpha + 4.013539 \Nf 
- 22.607687 ] a^2 \,+\, O(a^3)
\end{eqnarray}
where in keeping with observations in previous work in the MOM schemes the same
conversion function emerges in each scheme. Equipped with each conversion 
function then the renormalization group relation between the operator anomalous
dimensions is given formally by
\begin{eqnarray}
\gamma^{\MOMis}_{{\bar{\psi}\psi}} \left(a_{\MOMis}, \alpha_{\MOMis}\right) &=&
\left[ \frac{}{} \gamma_{{\bar{\psi}\psi}} (a) ~-~ 
\beta(a) \frac{\partial ~}{\partial a} \ln 
C^{\MOMis}_{\bar{\psi}\psi}(a,\alpha)
\right. \nonumber \\
&& \left. ~-~ \alpha \gamma_\alpha (a,\alpha)
\frac{\partial ~}{\partial \alpha} \ln C^{\MOMis}_{\bar{\psi}\psi} (a,\alpha)
\right]_{ \MSbars \rightarrow \MOMis } ~.
\label{anomcon}
\end{eqnarray}
Here the subscript mapping indicates that after the quantity in square brackets
has been determined then that expression which is in $\MSbar$ variables is
mapped to $\MOMi$ variables consistent with the arguments of the function on
the left hand side. In (\ref{anomcon}) the $\MSbar$ quark mass anomalous
dimension only depends on the coupling constant since that scheme is a mass
independent one and it is known, \cite{68}, that in that case the anomalous
dimension does not depend on $\alpha$. By contrast, the MOM scheme is a mass
dependent scheme and therefore anomalous dimensions of gauge invariant
operators will depend on the gauge parameter. We again note that for our 
purposes that although we include the gauge parameter throughout, our focus 
in analysing the critical exponents here will be solely on the Landau gauge.  

Having described the method we have used to evaluate the quark mass anomalous
dimension in each of the three MOM schemes we now record their explicit values 
for the Landau gauge. We have
\begin{eqnarray}
\gamma^{\MOMqs}_{\bar{\psi}\psi}(a,0) &=& -~ 3 C_F a \nonumber \\ 
&& + \left[ 
\left[
2
+ \frac{8}{9} \pi^2
- \frac{4}{3} \psi^\prime(\third)
\right] \Nf T_F C_F
+ \left[
- \frac{13}{4}
- \pi^2
+ \frac{3}{2} \psi^\prime(\third)
\right]
C_F C_A
\right. \nonumber \\
&& \left. ~~~
+ \left[
- \frac{27}{2}
- \frac{8}{9} \pi^2
+ \frac{4}{3} \psi^\prime(\third)
\right]
C_F^2
\right] a^2
\nonumber \\
&& + \left[ 
\left[
41
- \frac{20}{3} \zeta(3)
- \frac{16}{9} \pi^2
- \frac{8}{27} \pi^4
- 8 s_2(\pisix)
+ 16 s_2(\pitwo)
+ \frac{40}{3} s_3(\pisix)
- \frac{32}{3} s_3(\pitwo)
\right. \right. \nonumber \\
&& \left. \left. ~~~~
+ \frac{8}{3} \psi^\prime(\third)
+ \frac{16}{9} \psi^\prime(\third) \pi^2
- \frac{4}{3} \left( \psi^\prime(\third) \right)^2 
- \frac{1}{9} \psi^{\prime\prime\prime}(\third)
- \frac{1}{54} \ln^2(3) \sqrt{3} \pi
\right. \right. \nonumber \\
&& \left. \left. ~~~~
+ \frac{2}{9} \ln(3) \sqrt{3} \pi
+ \frac{29}{1458} \sqrt{3} \pi^3
\right]
\Nf T_F C_F C_A
\right. \nonumber \\
&& \left. ~~~
+ \left[
\frac{130}{3}
- \frac{32}{3} \zeta(3)
+ \frac{40}{9} \pi^2
- \frac{64}{81} \pi^4
+ 64 s_2(\pisix)
- 128 s_2(\pitwo)
- \frac{320}{3} s_3(\pisix)
\right. \right. \nonumber \\
&& \left. \left. ~~~~~~~
+ \frac{256}{3} s_3(\pitwo)
- \frac{20}{3} \psi^\prime(\third)
+ \frac{16}{9} \psi^\prime(\third) \pi^2
- \frac{4}{3} \left( \psi^\prime(\third) \right)^2 
+ \frac{2}{27} \psi^{\prime\prime\prime}(\third)
\right. \right. \nonumber \\
&& \left. \left. ~~~~~~~
+ \frac{4}{27} \ln^2(3) \sqrt{3} \pi
- \frac{16}{9} \ln(3) \sqrt{3} \pi
- \frac{116}{729} \sqrt{3} \pi^3
\right]
\Nf T_F C_F^2
- 8 \Nf^2 T_F^2 C_F
\right. \nonumber \\
&& \left. ~~~
+ \left[
- \frac{249}{4}
+ \frac{2503}{48} \zeta(3)
- \frac{1297}{72} \pi^2
- \frac{191}{486} \pi^4
+ \frac{347}{2} s_2(\pisix)
- 347 s_2(\pitwo)
\right. \right. \nonumber \\
&& \left. \left. ~~~~~~~
- \frac{1735}{6} s_3(\pisix)
+ \frac{694}{3} s_3(\pitwo)
+ \frac{1297}{48} \psi^\prime(\third)
+ \frac{175}{324} \psi^\prime(\third) \pi^2
- \frac{175}{432} \left( \psi^\prime(\third) \right)^2 
\right. \right. \nonumber \\
&& \left. \left. ~~~~~~~
+ \frac{23}{288} \psi^{\prime\prime\prime}(\third)
+ \frac{347}{864} \ln^2(3) \sqrt{3} \pi
- \frac{347}{72} \ln(3) \sqrt{3} \pi
- \frac{10063}{23328} \sqrt{3} \pi^3
\right]
C_F C_A^2
\right. \nonumber \\
&& \left. ~~~
+ \left[
- \frac{467}{12}
+ \frac{106}{3} \zeta(3)
+ \frac{515}{9} \pi^2
+ \frac{1216}{243} \pi^4
- 428 s_2(\pisix)
+ 856 s_2(\pitwo)
\right. \right. \nonumber \\
&& \left. \left. ~~~~~~~
+ \frac{2140}{3} s_3(\pisix)
- \frac{1712}{3} s_3(\pitwo)
- \frac{515}{6} \psi^\prime(\third)
- \frac{1192}{81} \psi^\prime(\third) \pi^2
+ \frac{298}{27} \left( \psi^\prime(\third) \right)^2 
\right. \right. \nonumber \\
&& \left. \left. ~~~~~~~
- \frac{1}{27} \psi^{\prime\prime\prime}(\third)
- \frac{107}{108} \ln^2(3) \sqrt{3} \pi
+ \frac{107}{9} \ln(3) \sqrt{3} \pi
+ \frac{3103}{2916} \sqrt{3} \pi^3
\right]
C_F^2 C_A
\right. \nonumber \\
&& \left. ~~~
+ \left[
- \frac{279}{2}
+ 56 \zeta(3)
- \frac{364}{9} \pi^2
+ \frac{176}{243} \pi^4
- 48 s_2(\pisix)
+ 96 s_2(\pitwo)
+ 80 s_3(\pisix)
\right. \right. \nonumber \\
&& \left. \left. ~~~~~~~
- 64 s_3(\pitwo)
+ \frac{182}{3} \psi^\prime(\third)
+ \frac{400}{81} \psi^\prime(\third) \pi^2
- \frac{100}{27} \left( \psi^\prime(\third) \right)^2 
- \frac{8}{9} \psi^{\prime\prime\prime}(\third)
\right. \right. \nonumber \\
&& \left. \left. ~~~~~~~
- \frac{1}{9} \ln^2(3) \sqrt{3} \pi
+ \frac{4}{3} \ln(3) \sqrt{3} \pi
+ \frac{29}{243} \sqrt{3} \pi^3
 \right]
C_F^3
\right] a^3 ~+~ O(a^4)
\end{eqnarray}
for the $\MOMq$ scheme and 
\begin{eqnarray}
\gamma^{\MOMgs}_{\bar{\psi}\psi}(a,0) &=& -~ 3 C_F a \nonumber \\ 
&& + \left[ 
\left[
\frac{2}{3}
+ \frac{88}{27} \pi^2
- \frac{44}{9} \psi^\prime(\third)
\right]
\Nf T_F C_F
+ \left[ 
- \frac{53}{6}
- \frac{89}{27} \pi^2
+ \frac{89}{18} \psi^\prime(\third)
\right]
C_F C_A
\right. \nonumber \\
&& \left. ~~~
- \frac{3}{2} C_F^2
\right] a^2
\nonumber \\
&&
+ \left[ 
\left[ 
\frac{2369}{54}
- \frac{128}{3} \zeta(3)
+ \frac{226}{243} \pi^2
+ \frac{12688}{2187} \pi^4
- \frac{377}{243\sqrt{3}} \pi^3
- \frac{52}{3\sqrt{3}} \ln(3) \pi
\right. \right. \nonumber \\
&& \left. \left. ~~~~
+ \frac{13}{9\sqrt{3}} \ln^2(3) \pi
+ 208 s_2(\pisix)
- 416 s_2(\pitwo)
- \frac{1040}{3} s_3(\pisix)
+ \frac{832}{3} s_3(\pitwo)
\right. \right. \nonumber \\
&& \left. \left. ~~~~
- \frac{113}{81} \psi^\prime(\third)
- \frac{15280}{729} \psi^\prime(\third) \pi^2
+ \frac{3820}{243} \left( \psi^\prime(\third) \right)^2 
+ \frac{4}{9} \psi^{\prime\prime\prime}(\third)
\right]
\Nf T_F C_F C_A
\right. \nonumber \\
&& \left. ~~~
+ \left[ 
18
- \frac{32}{3} \zeta(3)
+ \frac{104}{27} \pi^2
- \frac{320}{243} \pi^4
- \frac{116}{243\sqrt{3}} \pi^3
- \frac{16}{3\sqrt{3}} \ln(3) \pi
\right. \right. \nonumber \\
&& \left. \left. ~~~~~~~
+ \frac{4}{9\sqrt{3}} \ln^2(3) \pi
+ 64 s_2(\pisix)
- 128 s_2(\pitwo)
- \frac{320}{3} s_3(\pisix)
+ \frac{256}{3} s_3(\pitwo)
\right. \right. \nonumber \\
&& \left. \left. ~~~~~~~
- \frac{52}{9} \psi^\prime(\third)
+ \frac{272}{81} \psi^\prime(\third) \pi^2
- \frac{68}{27} \left( \psi^\prime(\third) \right)^2 
+ \frac{2}{27} \psi^{\prime\prime\prime}(\third)
\right]
\Nf T_F C_F^2
\right. \nonumber \\
&& \left. ~~~
+ \left[ 
- \frac{196}{27}
+ \frac{320}{243} \pi^2
- \frac{10240}{2187} \pi^4
- \frac{160}{81} \psi^\prime(\third)
+ \frac{10240}{729} \psi^\prime(\third) \pi^2
\right. \right. \nonumber \\
&& \left. \left. ~~~~~~~
- \frac{2560}{243} \left( \psi^\prime(\third) \right)^2 
\right]
\Nf^2 T_F^2 C_F
\right. \nonumber \\
&& \left. ~~~
+ \left[ 
- \frac{220159}{1728}
+ \frac{6367}{48} \zeta(3)
+ \frac{1643}{243} \pi^2
- \frac{9779}{17496} \pi^4
+ \frac{12499}{3888\sqrt{3}} \pi^3
\right. \right. \nonumber \\
&& \left. \left. ~~~~~~~
+ \frac{431}{12\sqrt{3}} \ln(3) \pi
- \frac{431}{144\sqrt{3}} \ln^2(3) \pi
- 431 s_2(\pisix)
+ 862 s_2(\pitwo)
\right. \right. \nonumber \\
&& \left. \left. ~~~~~~~
+ \frac{2155}{3} s_3(\pisix)
- \frac{1724}{3} s_3(\pitwo)
- \frac{1643}{162} \psi^\prime(\third)
+ \frac{22183}{2916} \psi^\prime(\third) \pi^2
\right. \right. \nonumber \\
&& \left. \left. ~~~~~~~
- \frac{22183}{3888} \left( \psi^\prime(\third) \right)^2 
- \frac{427}{576} \psi^{\prime\prime\prime}(\third)
\right]
C_F C_A^2
\right. \nonumber \\
&& \left. ~~~
+ \left[ 
13
+ \frac{88}{3} \zeta(3)
+ \frac{593}{27} \pi^2
+ \frac{880}{243} \pi^4
+ \frac{319}{243\sqrt{3}} \pi^3
+ \frac{44}{3\sqrt{3}} \ln(3) \pi
\right. \right. \nonumber \\
&& \left. \left. ~~~~~~~
- \frac{11}{9\sqrt{3}} \ln^2(3) \pi
- 176 s_2(\pisix)
+ 352 s_2(\pitwo)
+ \frac{880}{3} s_3(\pisix)
- \frac{704}{3} s_3(\pitwo)
\right. \right. \nonumber \\
&& \left. \left. ~~~~~~~
- \frac{593}{18} \psi^\prime(\third)
- \frac{748}{81} \psi^\prime(\third) \pi^2
+ \frac{187}{27} \left( \psi^\prime(\third) \right)^2 
- \frac{11}{54} \psi^{\prime\prime\prime}(\third)
\right]
C_F^2 C_A
\right. \nonumber \\
&& \left. ~~~
- \frac{129}{2} C_F^3 
\right] a^3 ~+~ O(a^4) \\
\gamma^{\MOMhs}_{\bar{\psi}\psi}(a,0) &=& -~ 3 C_F a \nonumber \\ 
&& + \left[
\left[
2
+ \frac{8}{9} \pi^2
- \frac{4}{3} \psi^\prime(\third)
\right]
\Nf T_F C_F
+ \left[
- \frac{55}{4}
- \frac{49}{18} \pi^2
+ \frac{49}{12} \psi^\prime(\third)
\right]
C_F C_A
\right. \nonumber \\
&& \left. ~~~
- \frac{3}{2} C_F^2
\right] a^2
\nonumber \\
&&
+ \left[
\left[
\frac{157}{2}
- \frac{32}{3} \zeta(3)
+ \frac{313}{27} \pi^2
+ \frac{104}{243} \pi^4
- \frac{29}{243\sqrt{3}} \pi^3
- \frac{4}{3\sqrt{3}} \ln(3) \pi
\right. \right. \nonumber \\
&& \left. \left. ~~~~
+ \frac{1}{9\sqrt{3}} \ln^2(3) \pi
+ 16 s_2(\pisix)
- 32 s_2(\pitwo)
- \frac{80}{3} s_3(\pisix)
+ \frac{64}{3} s_3(\pitwo)
- \frac{313}{18} \psi^\prime(\third)
\right. \right. \nonumber \\
&& \left. \left. ~~~~
- \frac{104}{81} \psi^\prime(\third) \pi^2
+ \frac{26}{27} \left( \psi^\prime(\third) \right)^2 
\right]
\Nf T_F C_F C_A
\right. \nonumber \\
&& \left. ~~~
+ \left[
\frac{46}{3}
- \frac{32}{3} \zeta(3)
- \frac{152}{27} \pi^2
- \frac{320}{243} \pi^4
- \frac{116}{243\sqrt{3}} \pi^3
- \frac{16}{3\sqrt{3}} \ln(3) \pi
\right. \right. \nonumber \\
&& \left. \left. ~~~~~~~
+ \frac{4}{9\sqrt{3}} \ln^2(3) \pi
+ 64 s_2(\pisix)
- 128 s_2(\pitwo)
- \frac{320}{3} s_3(\pisix)
+ \frac{256}{3} s_3(\pitwo)
\right. \right. \nonumber \\
&& \left. \left. ~~~~~~~
+ \frac{76}{9} \psi^\prime(\third)
+ \frac{272}{81} \psi^\prime(\third) \pi^2
- \frac{68}{27} \left( \psi^\prime(\third) \right)^2 
+ \frac{2}{27} \psi^{\prime\prime\prime}(\third)
\right]
\Nf T_F C_F^2
\right. \nonumber \\
&& \left. ~~~
+ \left[
- \frac{1419}{8}
+ \frac{3757}{48} \zeta(3)
- \frac{3383}{108} \pi^2
- \frac{469}{486} \pi^4
+ \frac{1015}{3888\sqrt{3}} \pi^3
+ \frac{35}{12\sqrt{3}} \ln(3) \pi
\right. \right. \nonumber \\
&& \left. \left. ~~~~~~~
- \frac{35}{144\sqrt{3}} \ln^2(3) \pi
- 35 s_2(\pisix)
+ 70 s_2(\pitwo)
+ \frac{175}{3} s_3(\pisix)
- \frac{140}{3} s_3(\pitwo)
\right. \right. \nonumber \\
&& \left. \left. ~~~~~~~
+ \frac{3383}{72} \psi^\prime(\third)
+ \frac{4751}{1296} \psi^\prime(\third) \pi^2
- \frac{4751}{1728} \left( \psi^\prime(\third) \right)^2 
- \frac{37}{384} \psi^{\prime\prime\prime}(\third)
\right]
C_F C_A^2
\right. \nonumber \\
&& \left. ~~~
+ \left[
\frac{97}{12}
+ \frac{88}{3} \zeta(3)
+ \frac{1217}{54} \pi^2
+ \frac{880}{243} \pi^4
+ \frac{319}{243\sqrt{3}} \pi^3
+ \frac{44}{3\sqrt{3}} \ln(3) \pi
\right. \right. \nonumber \\
&& \left. \left. ~~~~~~~
- \frac{11}{9\sqrt{3}} \ln^2(3) \pi
- 176 s_2(\pisix)
+ 352 s_2(\pitwo)
+ \frac{880}{3} s_3(\pisix)
- \frac{704}{3} s_3(\pitwo)
\right. \right. \nonumber \\
&& \left. \left. ~~~~~~~
- \frac{1217}{36} \psi^\prime(\third)
- \frac{748}{81} \psi^\prime(\third) \pi^2
+ \frac{187}{27} \left( \psi^\prime(\third) \right)^2 
- \frac{11}{54} \psi^{\prime\prime\prime}(\third)
\right]
C_F^2 C_A
\right. \nonumber \\
&& \left. ~~~
- 8 \Nf^2 T_F^2 C_F
- \frac{129}{2} C_F^3
\right] a^3 ~+~ O(a^4) 
\label{quarkmassmom}
\end{eqnarray}
for $\MOMg$ and $\MOMh$ respectively. Here $C_F$ and $C_A$ are the usual rank 
$2$ Casimirs in the fundamental and adjoint representations respectively which 
have dimensions $\NF$ and $\NA$. The Dynkin index is $T_F$ and $\Nf$ is the 
number of massless quark flavours. Various numbers arise through the values of 
the underlying masters. The function $\psi(z)$ is the Euler $\psi$-function and
$\zeta(z)$ is the Riemann zeta function. Various specific values of the 
polylogarithm function, $\mbox{Li}_n(z)$, occur which are defined by 
\begin{equation}
s_n(z) ~=~ \frac{1}{\sqrt{3}} \Im \left[ \mbox{Li}_n \left(
\frac{e^{iz}}{\sqrt{3}} \right) \right] ~.
\end{equation} 
To assist the evaluation of the quark mass anomalous dimensions numerically we
note
\begin{eqnarray}
\zeta(3) &=& 1.20205690 ~,~
\psi^\prime ( \third ) ~=~ 10.09559713 ~,~
\psi^{\prime\prime\prime} ( \third ) ~=~ 488.1838167 ~,~
s_2 ( \pitwo ) ~=~ 0.32225882 
\nonumber \\
s_2 ( \pisix ) &=& 0.22459602 ~,~
s_3 ( \pitwo ) ~=~ 0.32948320 ~,~
s_3 ( \pisix ) ~=~ 0.19259341 ~.
\end{eqnarray}
Throughout we use $a$~$=$~$g^2/(16\pi^2)$ as the coupling constant in keeping
with conventions used in previous articles. Equipped with these expressions we
are now in a position to analyse their values at the Banks-Zaks fixed point.

\sect{Fixed points.}

Before carrying out that analysis we concentrate in this section on various
aspects of fixed points in the QCD $\beta$-function. For the moment we review
the situation in the $\MSbar$ scheme partly as this forms the discussion but
partly as this was the scheme in which the Banks-Zaks fixed point was explored 
initially, \cite{3,5}. Although the four loop $\MSbar$ $\beta$-function is 
available, and will be used later, for the moment the three loop result of
\cite{19} is sufficient for the present discussion and is  
\begin{eqnarray}
\beta^{\MSbars}(a) &=& -~ \left[ \frac{11}{3} C_A - \frac{4}{3} T_F \Nf 
\right] a^2 ~-~
\left[ \frac{34}{3} C_A^2 - 4 C_F T_F \Nf - \frac{20}{3} C_A T_F \Nf \right]
a^3 \nonumber \\
&& +~ \left[ 2830 C_A^2 T_F \Nf - 2857 C_A^3 + 1230 C_A C_F T_F \Nf
- 316 C_A T_F^2 \Nf^2 \right. \nonumber \\
&& \left. ~~~~~-~ 108 C_F^2 T_F \Nf - 264 C_F T_F^2 \Nf^2 \right]
\frac{a^4}{54} ~+~ O(a^5)
\label{betams}
\end{eqnarray}
in four dimensions. The observation of \cite{5} was basically that at two loops
for a range of values of $\Nf$ the $\beta$-function has a non-trivial zero 
which we formally denote by $a_2$. This arises when the first term of 
(\ref{betams}) is negative and when the second term is positive. For a 
sufficiently large number of massless quarks asymptotic freedom is lost and the
theory becomes like QED. When a real positive non-trivial solution exists then 
this is termed the Banks-Zaks fixed point. As it occurs for that part of the 
$\beta$-function which is scheme independent then it should be a universal 
property of the theory. However, with the inclusion of higher order terms in 
$\beta(a)$ not only will the location of the fixed point be refined but its 
specific value will be scheme dependent. So, for example, denoting the three 
and four loop Banks-Zaks fixed points by $a_3$ and $a_4$ respectively, then 
these would depend on the renormalization scheme which $\beta(a)$ was expressed
in. In the case of $a_4$ and higher fixed points there could be more than one 
non-trivial root of $\beta(a)$~$=$~$0$. The Banks-Zaks one is always regarded 
as the one closest to the origin. Some remarks are apt on the scheme dependence
of the range of $\Nf$ for which the conformal window exists. From
(\ref{betams}) the upper limit of the range is determined by the one loop
coefficient while the two loop term gives the lower limit. For $SU(3)$ the
lower limit is $\Nf$~$=$~$9$. If the Banks-Zaks fixed point is related to the 
breaking of chiral symmetry then it would appear to be ruled out in this 
scenario. However, the conformal window discussed so far is deduced from a 
perturbative analysis and, moreover, the value of the critical coupling for low 
values of $\Nf$ in the window are outside the region of perturbative 
credibility. Indeed it could be the case that when lower values of $\Nf$ are 
analysed non-perturbatively then the lower boundary of the window could be 
reduced. A second aspect of the lower end is that it derives from the two loop 
coefficient of $\beta(g)$. In mass independent renormalization schemes where 
there is a single coupling constant this term is scheme independent, \cite{68}.
However, in MOM schemes with a non-zero $\alpha$ this term is both $\alpha$ and
scheme dependent. In the Landau gauge the two loop terms of each of the MOM 
$\beta$-functions reduce to the same value as (\ref{betams}). This may not be
the case in other gauges such as a non-linear gauge. For instance, in 
\cite{69} the two loop renormalization group functions have been deduced in the
corresponding MOM schemes for the maximal abelian gauge, \cite{70,71,72}. From 
those results using the two loop term in the corresponding $\beta$-functions 
the lower bound of the conformal window for two of the schemes drops to 
$\Nf$~$=$~$8$. Again this is a perturbative observation in a region where the 
location of the fixed point lies outside the range of validity of perturbation 
theory. So it does not imply that the lower limit of the conformal window is 
lower ahead of a full non-perturbative analysis. Though the lattice study of 
\cite{8} has provided evidence that the low end of the window can accommodate 
this value. Not only has the location of the fixed point been studied in 
various schemes in \cite{17,31,32} but the values of the quark mass anomalous 
dimension at the Banks-Zaks fixed point have been estimated in the same 
schemes. As these critical exponents are renormalization group invariant it 
should be the case that with sufficiently high accuracy the scheme dependence 
evident in lower loop estimates should wash out. That has been observed in
\cite{17} for certain values of $\Nf$ in the window where the Banks-Zaks fixed 
point exists. This is invariably for large values of $\Nf$ close to the upper 
boundary. For lower values of $\Nf$ the value of $a_n$ ceases to be small and 
so estimates of critical exponents would be outside the perturbative region. 
For $\Nf$ in the intermediate part of the range it may be the case that the 
higher order corrections restore $a_n$ to perturbative reliability. In addition
certain schemes may remain within the perturbative region better than others. 
This is one question we aim to analyse. 

In respect of these points it is worth noting the structure of the MOM scheme 
$\beta$-functions we will use to deduce the quark mass critical exponents at
$a_n$. As these expressions have been given elsewhere, \cite{44,45}, and are 
equally as cumbersome as (\ref{quarkmassmom}), we record the expression for the
$\MOMh$ scheme in the Landau gauge as it is the more compact of the three. It 
is, \cite{44,45},
\begin{eqnarray}
\beta^{\MOMhs}(a,0) &=& -~ \left[ \frac{11}{3} C_A - \frac{4}{3} 
T_F \Nf \right] a^2 ~-~ \left[ \frac{34}{3} C_A^2 - 4 C_F T_F \Nf 
- \frac{20}{3} C_A T_F \Nf \right] a^3 \nonumber \\
&& +~ \left[ \left[
18817920
+ 103680 \pi^2
- 16422912 \zeta(3)
- 155520 \psi^\prime(\third)
\right]
\Nf T_F C_A C_F
\right. \nonumber \\
&& \left. ~~~~~
+ \left[
29167776
+ 3729024 \pi^2
+ 29568 \pi^4
+ 11562912 \zeta(3)
- 5593536 \psi^\prime(\third)
\right. \right. \nonumber \\
&& \left. \left. ~~~~~~~~~
+ 7200 \pi^2 \psi^\prime(\third) 
- 5400 (\psi^\prime(\third))^2 
- 11988 \psi^{\prime\prime\prime}(\third)
- 31726080 s_2( \pisix)
\right. \right. \nonumber \\
&& \left. \left. ~~~~~~~~~
+ 63452160 s_2( \pitwo)
+ 52876800 s_3( \pisix)
- 42301440 s_3( \pitwo)
+ 78880 \pi^3 \sqrt{3}
\right. \right. \nonumber \\
&& \left. \left. ~~~~~~~~~
+ 881280 \ln(3) \pi \sqrt{3}
- 73440 \ln^2(3) \pi \sqrt{3}
\right]
\Nf T_F C_A^2 \right. \nonumber \\
&& \left. ~~~~~
+ \left[
- 4105728
- 705024 \pi^2
- 3981312 \zeta(3)
+ 1057536 \psi^\prime(\third)
+ 5971968 s_2( \pisix)
\right. \right. \nonumber \\
&& \left. \left. ~~~~~~~~~
- 11943936 s_2( \pitwo)
- 9953280 s_3( \pisix)
+ 7962624 s_3( \pitwo)
- 14848 \pi^3 \sqrt{3}
\right. \right. \nonumber \\
&& \left. \left. ~~~~~~~~~
- 165888 \ln(3) \pi \sqrt{3}
+ 13824 \ln^2(3) \pi \sqrt{3}
\right]
\Nf^2 T_F^2 C_A
\right. \nonumber \\
&& \left. ~~~~~
+ \left[
- 5723136
+ 5971968 \zeta(3)
\right]
\Nf^2 T_F^2 C_F
- 559872 \Nf T_F C_F^2
\right. \nonumber \\
&& \left. ~~~~~
+ \left[
- 35200008
- 4741632 \pi^2
- 81312 \pi^4
- 1689336 \zeta(3)
+ 7112448 \psi^\prime(\third)
\right. \right. \nonumber \\
&& \left. \left. ~~~~~~~~~
- 19800 \pi^2 \psi^\prime(\third) 
+ 14850 (\psi^\prime(\third))^2 
+ 32967 \psi^{\prime\prime\prime}(\third)
+ 42083712 s_2( \pisix)
\right. \right. \nonumber \\
&& \left. \left. ~~~~~~~~~
- 84167424 s_2( \pitwo)
- 70139520 s_3( \pisix)
+ 56111616 s_3( \pitwo)
- 104632 \pi^3 \sqrt{3}
\right. \right. \nonumber \\
&& \left. \left. ~~~~~~~~~
- 1168992 \ln(3) \pi \sqrt{3}
+ 97416 \ln^2(3) \pi \sqrt{3}
\right]
C_A^3
\right] \frac{a^4}{279936} ~+~ O(a^5)
\label{betamomc}
\end{eqnarray}
at three loops. Like (\ref{quarkmassmom}) at three loops the presence of the
underlying symmetric point masters are evident. We note that we are effectively
quoting the full expression given in \cite{45} but with a modification. In the
three loop term of equation (5.28) in \cite{45} an additional numerical object,
$\Sigma$, was present which was a combination of harmonic polylogarithms. When 
\cite{45} appeared it was not apparent that this was not an independent 
quantity and has since been shown to correspond to, \cite{73},  
\begin{equation}
\Sigma ~=~ \frac{1}{36} \psi^{\prime\prime\prime}\left( \frac{1}{3} \right)
- \frac{2\pi^4}{27}
\label{sigmacomb}
\end{equation}
in the notation of the previous renormalization group equations. We have
substituted (\ref{sigmacomb}) in the original expression of \cite{45} for
consistency here. In comparing (\ref{betams}) and (\ref{betamomc}) one can see 
that there is a structural question to be addressed. If when one computes the 
critical exponent for, say, the quark mass anomalous dimension in $\MSbar$ and 
$\MOMh$ at the Banks-Zaks fixed point then both expressions ought to be the 
same. This is because ultimately the critical exponent is a physical quantity 
and hence a renormalization group invariant. It is independent of the 
renormalization scheme in which it is determined. However, given the form of 
both $\beta$-functions this cannot be the case. Indeed this is one of the 
motivations for examining the critical exponents at the Banks-Zaks fixed point 
in MOM schemes. These are clearly in a different class from the point of view 
of the numerology when compared with the schemes analysed in \cite{17} which 
were $\MSbar$, RI${}^\prime$ and $\mMOM$. The coefficients appearing in the 
renormalization group functions of these three schemes are from the set
\begin{equation}
\left\{ {\mathbb Q}, \pi^2, \zeta(3), \zeta(5) \right\}
\label{msbasis}
\end{equation}
to four loops. By contrast the basis for the MOM scheme coefficients to three 
loops is
\begin{equation}
\left\{ {\mathbb Q}, \pi^2, \zeta(3), \zeta(4),
\psi^\prime( \third ),
\psi^{\prime\prime\prime}( \third ),
s_2( \pitwo ),
s_2( \pisix ),
s_3( \pitwo ),
s_3( \pisix ),
\frac{\ln^2(3) \pi}{\sqrt{3}},
\frac{\ln(3) \pi}{\sqrt{3}},
\frac{\pi^3}{\sqrt{3}}
\right\} ~. 
\label{mombasis}
\end{equation}
The aim would be to see if the numerical values for the exponents in various
schemes show the consistency which would indicate renormalization group
invariance. 

This problem can also be illustrated in the context of another fixed point 
which is present in the QCD $\beta$-function but is usually discussed in the 
context of scalar field theories. It is the Wilson-Fisher fixed point,
\cite{74,75,76,77}, which occurs in the $d$-dimensional $\beta$-function. For 
QCD the latter, which will be denoted by $\beta^{\MSbars}_d(a)$ in $\MSbar$, is
related to (\ref{betams}) by
\begin{equation}
\beta^{\MSbars}_d(a) ~=~ \half (d-4) ~+~ \beta^{\MSbars}(a) ~.
\label{betamsd}
\end{equation}
Irrespective of whether there is a Banks-Zaks fixed point or not in QCD, there 
will be Wilson-Fisher fixed point in $d$~$<$~$4$ dimensions when the one loop 
term of $\beta(a)$ is {\em positive}. In scalar theories this fixed point has 
proved useful in obtaining estimates for critical exponents in {\em three} 
dimensions through, for example, resummation techniques. However, such critical 
exponents are also renormalization group invariant and therefore the explicit 
expressions should be equivalent. It has been possible to check this for 
certain scalar theories, \cite{78}. The same analysis can be studied here with 
(\ref{betamomc}) but the numeric structure of the renormalization group 
functions would appear to suggest otherwise given the number bases indicated 
above. This is not the case due to a subtle feature which is absent in 
(\ref{betamomc}). We have correctly introduced the concept of the Wilson-Fisher
fixed point for the $\MSbar$ scheme. For the $\MOMh$ scheme the situation is 
completely parallel except that one cannot merely replace the scheme label in 
(\ref{betamsd}) by the scheme label for $\MOMh$. This is because the MOM 
schemes are defined by the inclusion of finite parts in the renormalization 
constants. In the derivation of (\ref{betamomc}) the final step is to set 
$\epsilon$~$=$~$0$ whence (\ref{betamomc}) emerges. However, the $d$-dimension 
$\MOMh$ $\beta$-function has $\epsilon$-dependent coefficients unlike the 
$\MSbar$ scheme. In the latter scheme there are no finite contributions to the 
renormalization constants. So to check that the $\MOMh$ Wilson-Fisher fixed 
point actually delivers expressions for critical exponents which are equivalent
for different schemes one has to perform the analysis fully in $d$-dimensions. 
As the MOM renormalization group functions have not been recorded in 
$d$-dimensions for each of the three MOM schemes of Celmaster and Gonsalves we 
do so here for an interested reader. Though to save space we record the 
expressions numerically and provide the full analytic $d$-dimensional 
expressions in the accompanying data file. In that file the three loop terms in
the MOM renormalization group functions are not provided as the $\epsilon$
dependent terms at that loop order are derived from the {\em finite} parts of
the three loop renormalization constants. These are not known at present. We do
include the three loop coefficients in the data file for the quark mass
operator in the MOM schemes as they appear here for the first time. So, for 
instance, in the Landau gauge for the colour group $SU(3)$ we have  
\begin{eqnarray}
\beta^{\mbox{$\MOMqs$}}(a,0) &=&
-~ \epsilon a
+ \left[ 0.666667 \Nf - 11.000000 + 1.111111 \Nf \epsilon 
- 16.715775 \epsilon \right] a^2 \nonumber \\
&& +~ \left[ 12.666667 \Nf - 102.000000 + 91.930102 \epsilon \Nf - 385.483952 
\epsilon \right] a^3 ~+~ O(a^3)
\nonumber \\
\gamma_A^{\mbox{$\MOMqs$}}(a,0) &=&
\left[ 0.666667 \Nf - 6.500000 + 1.111111 \Nf \epsilon 
- 8.083333 \epsilon \right] a
\nonumber \\
&& +~ \left[ 9.411706 \Nf - 46.639132 + 62.308328 \Nf \epsilon 
- 311.747527 \epsilon \right] a^2 ~+~ O(a^3)
\nonumber \\
\gamma_\psi^{\mbox{$\MOMqs$}}(a,0) &=&
\left[ -~ 1.333333 \Nf + 22.333333 - 4.666667 \Nf \epsilon + 50.928412 \epsilon 
\right] a^2 ~+~ O(a^3)
\nonumber \\
\gamma_c^{\mbox{$\MOMqs$}}(a,0) &=&
\left[ -~ 2.250000 - 3.000000 \epsilon \right] a
\nonumber \\
&& +~ \left[ 0.750000 \Nf - 13.202007 + 8.541667 \Nf \epsilon 
- 102.724216 \epsilon \right] a^2 ~+~ O(a^3)
\nonumber \\
\gamma_{\bar{\psi}\psi}^{\mbox{$\MOMqs$}}(a,0) &=&
\left[ -~ 4.000000 - 0.645519 \epsilon \right] a
\nonumber \\
&& +~ \left[ -~ 1.791876 \Nf - 7.570942 + 7.309836 \Nf \epsilon 
- 34.841722 \epsilon \right] a^2 ~+~ O(a^3)
\end{eqnarray}
for the $\MOMq$ scheme and 
\begin{eqnarray}
\beta^{\mbox{$\MOMgs$}}(a,0) &=&
-~ \epsilon a + \left[ 0.666667 \Nf - 11.000000
+ 3.416806 \Nf \epsilon - 26.492489 \epsilon \right] a^2 \nonumber \\
&& +~ \left[ 12.666667 \Nf - 102.000000 + 7.974346 \Nf^2 \epsilon 
+ 42.091196 \Nf \epsilon \right. \nonumber \\
&& \left. ~~~~~ -~ 517.221499 \epsilon \right] a^3 ~+~ O(a^4) 
\nonumber \\
\gamma_A^{\mbox{$\MOMgs$}}(a,0) &=&
\left[ 0.666667 \Nf - 6.500000 + 1.111111 \Nf \epsilon 
- 8.083333 \epsilon \right] a \nonumber \\
&& +~ \left[ 1.537130 \Nf^2 - 12.093123 \Nf + 16.909511 
+ 2.561884 \Nf^2 \epsilon 
\right. \nonumber \\
&& \left. ~~~~~ +~ 32.807608 \Nf \epsilon 
- 232.719087 \epsilon \right] a^2 ~+~ O(a^3) \nonumber \\
\gamma_\psi^{\mbox{$\MOMgs$}}(a,0) &=&
\left[ -~ 1.333333 \Nf + 22.333333 - 4.666667 \Nf \epsilon 
+ 50.928412 \epsilon \right] a^2 ~+~ O(a^3) \nonumber \\
\gamma_c^{\mbox{$\MOMgs$}}(a,0) &=&
\left[ -~ 2.250000 - 3.000000 \epsilon \right] a \nonumber \\
&& +~ \left[ -~ 4.437814 \Nf + 8.795600 + 1.624581 \Nf \epsilon 
- 73.394073 \epsilon \right] a^2 ~+~ O(a^3) \nonumber \\
\gamma_{\bar{\psi}\psi}^{\mbox{$\MOMgs$}}(a,0) &=&
\left[ -~ 4.000000 - 0.645519 \epsilon \right] a \nonumber \\
&& +~ \left[ -~ 11.014658 \Nf + 31.535915 + 5.821466 \Nf \epsilon 
- 28.530668 \epsilon \right] a^2 \nonumber \\
&& +~ O(a^3) \nonumber \\
\beta^{\mbox{$\MOMhs$}}(a,0) &=&
-~ \epsilon a + \left[ 0.666667 \Nf - 11.000000 + 1.111111 \Nf \epsilon 
- 18.548275 \epsilon \right] a^2 \nonumber \\
&& +~ \left[ 12.666667 \Nf - 102.000000 + 88.675121 \Nf \epsilon 
- 595.803097 \epsilon \right] a^3 ~+~ O(a^4) \nonumber \\
\gamma_A^{\mbox{$\MOMhs$}}(a,0) &=&
\left[ 0.666667 \Nf - 6.500000 + 1.111111 \Nf \epsilon - 8.083333 \epsilon
\right] a \nonumber \\
&& +~ \left[ 8.190039 \Nf - 34.727877 + 60.272216 \Nf \epsilon 
- 296.934812 \epsilon \right] a^2 ~+~ O(a^3) \nonumber \\
\gamma_\psi^{\mbox{$\MOMhs$}}(a,0) &=&
\left[ -~ 1.333333 \Nf + 22.333333 - 4.666667 \Nf \epsilon + 50.928412 \epsilon
\right] a^2 ~+~ O(a^3) \nonumber \\
\gamma_c^{\mbox{$\MOMhs$}}(a,0) &=&
\left[ -~ 2.250000 - 3.000000 \epsilon \right] a \nonumber \\
&& +~ \left[ 0.750000 \Nf - 9.078880 + 8.541667 \Nf \epsilon 
- 97.226714 \epsilon \right] a^2 ~+~ O(a^3) \nonumber \\
\gamma_{\bar{\psi}\psi}^{\mbox{$\MOMhs$}}(a,0) &=&
\left[ -~ 4.000000 - 0.645519 \epsilon \right] a \nonumber \\
&& +~ \left[ - 1.791876 \Nf - 0.240939 + 7.309836 \Nf \epsilon 
- 33.658808 \epsilon \right] a^2 ~+~ O(a^3)
\end{eqnarray}
for the other two MOM schemes. Of course this procedure can be reverse
engineered if one knew the $\beta$-function in one scheme to $L$ loops and to
$(L-1)$ in another scheme. In this instance some information on the latter
$\beta$-function can be adduced about the $L$-loop term from the
renormalization group invariance of the underlying critical exponent. Though
this is essentially reflective of the use of the conversion functions to 
establish the anomalous dimensions at the next order in a scheme in the context
we used earlier. Finally, we note that we have checked that the {\em same}
critical exponents emerge at the Wilson-Fisher fixed point in the MOM schemes
as in $\MSbar$ to $O(\epsilon^3)$ as expected. The situation for the 
Banks-Zaks fixed point is not the same primarily because it is a purely
{\em four} dimensional fixed point. Clearly the MOM critical exponents at the 
Banks-Zaks fixed point will involve the numbers in the basis (\ref{mombasis}) 
in contrast to the basis (\ref{msbasis}) for the schemes studied in \cite{17}. 
However, it is also partly due to the fact that when we compute the estimates 
of the quark mass critical exponent, for instance, we are endeavouring to use 
the perturbation theory of QCD which is valid in a region near the origin. 
Provided one is within the perturbative region of that theory then information 
about the exponents of the theory underpinning the Banks-Zaks fixed point can 
be obtained and should be comparable across schemes. However, to truly 
understand the renormalization group invariance via a scheme analysis of the 
Banks-Zaks fixed point one would first have to construct the quantum field 
theory which is in the same universality class and then renormalize it within 
the various schemes. That theory is not yet available as far as we are aware.

\sect{Results.}

Having discussed the nature of the two main critical points in the QCD 
renormalization group functions we turn now to the problem we will analyse with
them which is the evaluation of the quark mass anomalous dimension at the
Banks-Zaks fixed point. This will be carried out for a variety of colour groups
with the quarks in various representations. The usual case where the quarks
are in the fundamental representation will form the main part of the analysis.
However, for theories beyond the Standard Model, the analysis of \cite{17}
also included quarks in the adjoint representation as well as in the two-index
symmetric and antisymmetric representations for the RI${}^\prime$ and minimal 
MOM schemes. We will therefore provide results for these representations too in
order to have as large a picture as possible on where the convergence is best. 
For the explicit values of the various colour group Casimirs for these 
representations we refer the reader to Appendix B of \cite{17}. It is worth 
noting that the window for a Banks-Zaks fixed point depends on the particular 
representation and for some of these there is a much smaller range of $\Nf$ 
values for an infrared fixed point than that for quarks in the fundamental 
representation. Although the main interest is the quark mass anomalous 
dimension due to its relation to the conformal window, for a convergence 
analysis an equally useful critical exponent to analyse is that relating to the
critical slope of the $\beta$-function which is usually denoted by $\omega$. 
Its main role is as a measure of corrections to scaling. Therefore, we will be 
providing evaluations of this exponent for the same quark representations as 
the quark mass anomalous dimension. In order to present our results we need to 
introduce our notation. 

First, we formally define the Landau gauge $\beta$-function in the scheme 
${\cal S}$ by
\begin{equation}
\beta^{\cal S}(a,0) ~=~ \sum_{r=1}^\infty \beta_r^{\cal S} a^{r+1}
\end{equation}
and the $\beta$-function partial sums by  
\begin{equation}
\beta^{\cal S}_n(a,0) ~=~ \sum_{r=1}^n \beta_r^{\cal S} a^{r+1} ~. 
\end{equation}
Then for each scheme the Banks-Zaks fixed point $a_L$ at the $L$th loop order 
is defined as the first non-trivial zero of 
\begin{equation}
\beta^{\cal S}_L(a_L,0) ~=~ 0 ~.
\end{equation}
From this we define the critical exponent $\omega$ at the $L$th loop as
\begin{equation}
\omega_L ~=~ 2 \beta^\prime_L(a_L,0) ~. 
\end{equation}
For the Landau gauge quark mass anomalous dimension 
$\gamma^{\cal S}_{\bar{\psi}\psi}(a,0)$ we define the critical exponent by a 
similar process. We let the perturbative expression be
\begin{equation}
\gamma^{\cal S}_{\bar{\psi}\psi}(a,0) ~=~ \sum_{r=1}^\infty \gamma_r^{\cal S} 
a^r
\end{equation}
and then the corresponding partial sums are 
\begin{equation}
\gamma^{\cal S}_{\bar{\psi}\psi \, n}(a,0) ~=~ \sum_{r=1}^n \gamma_r^{\cal S} 
a^r ~.
\end{equation}
Denoting the quark mass anomalous dimension exponent by $\rho$ then its
evaluation at the $L$th loop fixed point is $\rho_L$ where
\begin{equation}
\rho_L ~=~ -~ 2 \gamma_{\bar{\psi}\psi \, L}(a_L,0) 
\end{equation}
for each scheme. The definition of $\rho$ coincides with that of \cite{17} so
that there is a direct comparison. However, given that we are defining the
$\beta$-function consistent with the conventions used in \cite{45}, the values
of the location of the fixed points differ by a factor of $4\pi$ from those of 
\cite{17}. Further, in presenting our results we use a similar form of tables 
but perform the evaluation to six decimal places. This is partly to compare the
convergence for certain cases. The format of the results tables parallels
\cite{17} in that we present the $\MSbar$ and $\mMOM$ results in a combined 
table since there are results to four loops for these two schemes. Subsequently
the results for the same quantity in the three MOM schemes are given. The order
within each choice of quark representation is fixed point location, $\omega$ 
and $\rho$. Though in one instance we include results for the 't Hooft scheme 
of \cite{79}. Briefly the renormalization group functions of this scheme are 
defined as that part which is renormalization scheme independent. For the 
$\beta$-function this is the two loop part and for the quark mass anomalous 
dimension it is the one loop term, \cite{70}. Our final general comment on the 
tables of results concerns the situation with the $\mMOM$ scheme. It transpires
that in \cite{41} there was an error in the derivation of the four loop quark 
mass anomalous dimension. Specifically the four loop term of the $\MSbar$ 
anomalous dimension was inadvertently subtracted in the corresponding 
derivation using (\ref{anomcon}). Therefore, the results of \cite{17} have been
corrected in an erratum using the erratum for \cite{41}. For completeness we 
also include the results for the correct version of the $\mMOM$ four loop quark
mass anomalous dimension to the accuracy we are working to.

We turn now to a discussion of the results in the individual Tables. For the
fundamental representation the fixed point locations are given in Tables $1$  
and $2$ and we make no comment on them as comparison between schemes of the
location is not fully meaningful. One role they play is to give an indication 
as to where the fixed point is becoming reasonably stable for certain values of
$\Nf$. Then one would hope that the corresponding critical exponents could be 
converging. For instance, from Table $1$ it would appear that for 
$\Nf$~$\geq$~$13$ the fixed point has reached a plateau for each scheme from 
the stability at three and four loops. It was noted in \cite{17} that the 
convergence is best at the upper end of the window for the infrared fixed 
point. This is because one is still in the region where the coupling constant 
has a small value. For smaller values of $\Nf$ the perturbative results do not 
appear to be reliable. Throughout our analysis we are broadly in agreement with
this point of view. As $\Nf$~$=$~$12$ is the value which is of intense interest
in the lattice community the perturbative results may not be competitive with 
that analysis. Given this we are not in a position to indicate whether the same
range of $\Nf$ values for perturbative reliability are valid in the MOM case as
we only have three loop results as is evident in Table $2$. In both instances, 
another way of examining convergence and the relevant $\Nf$ window is to 
examine the renormalization group invariant critical exponents. For $\omega$ 
these are given in Tables $3$ and $4$ for the five schemes we are interested 
in. In this and further remarks our discussion will always concentrate on the 
$\Nc$~$=$~$3$ case, unless otherwise indicated, due to the relation to QCD but 
for $\Nc$~$=$~$2$ and $3$ parallel remarks will apply but for different $\Nf$ 
values. From Tables $3$ and $4$ the three loop values of $\omega$ are all in 
accord for $\Nf$~$=$~$16$ as expected although the $\MOMg$ scheme is slightly 
lower. Indeed this is the case for lower values of $\Nf$. For instance, when 
$\Nf$~$=$~$13$ the four loop $\MSbar$ and $\mMOM$ values of $\omega$ are 
similar to the three loop ones of $\MOMq$ and $\MOMh$. However, when 
$\Nf$~$=$~$12$ this relative convergence is absent as expected. 

In general there is a parallel picture for the $\rho$ in Tables $5$ and $6$. 
Before concentrating on the five schemes we are interested in, in the former
Table we have included an additional column for the 't Hooft scheme. This was 
not needed for $\omega$ since the two loop $\MSbar$ column in Table $3$
corresponds to that scheme. However, for the quark mass anomalous dimension
case it appears evident that for a large range of $\Nf$ in the fixed point
window the estimates lie well away from those of our five schemes. This is not
surprising given the way the series is defined. Focusing now on our five 
schemes the three loop $\Nf$~$=$~$16$ values are comparable although again the
$\MOMg$ value appears to be the outlier here being on the higher side which is
also reflected at lower values of $\Nf$. At $\Nf$~$=$~$13$ the four loop 
$\MSbar$ and $\MOMq$ values are similar. The $\mMOM$ values are higher but
appear to be slowly decreasing. We note that with the previous wrong result the
$\mMOM$ four loop estimates were all higher than the three loop value. This is 
not the case now and is in fact reversed when the correct four loop 
expression is used. At $\Nf$~$=$~$12$ the situation is similar to $\omega$.
However, in this instance there are various lattice estimates for what we
have termed $\rho$. For instance, one particular analysis gives the value
of $0.235(15)$ from \cite{14} and another more recent study gives $0.235(46)$
from \cite{15}. In either case these values are lower than any of the three or 
four loop perturbative estimates and so we reinforce the observation of
\cite{17} that non-perturbative properties may be beginning to dominate the 
window at this point. One interesting feature of this $\Nf$ value is that if 
the lattice estimate is roughly correct the four loop $\MSbar$ value of $\rho$ 
is the closest. However, in terms of convergence the {\em three} loop $\MOMq$ 
and $\MOMh$ values are smaller than the corresponding three loop $\MSbar$ one. 
Hence one hope would be that a four loop analysis in these two schemes may
produce a better estimate in comparison to \cite{14,15} than the four loop 
$\MSbar$ one. In some sense since we are evaluating the quark mass anomalous 
dimension exponent it might be expected that the $\MOMq$ scheme would produce 
the more reliable value. That the $\MOMh$ value is competitive may seem 
surprising but given the similar structure of the Feynman graphs within the 
vertex functions defining each of the $\MOMq$ and $\MOMh$ schemes this would 
appear to be the main explanation. In each case one renormalizes the same 
number of graphs in the respective vertex renormalizations, \cite{45}, and the 
graphs are effectively the same structure topologically when examined in 
detail. 

Our final remark on the estimates of $\rho$ for $\Nf$~$=$~$12$ specifically 
concerns the use of the {\em five} loop quark mass anomalous dimension which 
was recently determined in \cite{50} in $\MSbar$ but specifically for 
$\Nc$~$=$~$3$. Although the five loop $\beta$-function is not available we
have carried out a tentative analysis using the expression given in \cite{50}.
The corresponding results are presented in Table $7$ where we have used an
additional notation, $\rho_{5l}$. This indicates the use of the five loop
$\MSbar$ quark mass anomalous dimension of \cite{50} but evaluated with the
corresponding values of the $l$-loop fixed points given in the $\MSbar$ columns
of Table $1$. The reason for using the three and four loop fixed point values 
is that if there is perturbative convergence it would be hoped that these would
bound the actual five loop value which is as yet unknown. As $a_4$~$>$~$a_3$ we
have assumed without any justification that there is such an alternating 
convergence. So if these are the bounding values the same reasoning would be
that $\rho_{53}$ and $\rho_{54}$ would bound the actual five loop value. This 
would appear to be the case for $\Nf$~$=$~$16$ as well as down to
$\Nf$~$=$~$13$ when comparing between schemes. If this reasoning applied to
$\Nf$~$=$~$12$ then the value of Table $7$ would appear to be significantly
different from the lattice estimates. By contrast another way of expressing
this is to determine the value of $a_5$ which would be required to give the
central value of \cite{14,15} of $\rho_5$~$=$~$0.235$. From the five loop 
expression of \cite{50} we would have to have $a_5$~$=$~$0.028376$ in our 
conventions which is significantly lower than the three and four loop values we
used to obtain the $\Nf$~$=$~$12$ estimates in Table $7$. In other words it is 
well inside the region where perturbation theory is valid and suggests that 
non-perturbative properties are the drive behind the two consistent lattice 
estimates. Such a large drop in the critical coupling value from successive 
loop orders is not seen in $\MSbar$ for $\Nc$~$=$~$3$ even for smaller values 
of $\Nf$. Finally, for fundamental quarks we note that the $SU(2)$ colour group
has been studied on the lattice for values of $\Nf$ in the range 
$6$~$\leq$~$\Nf$~$\leq$~$10$, \cite{10}. All our estimates for $\rho$ are in
good agreement with the $\Nf$~$=$~$10$ value of $0.08$ given in \cite{10}. This
is in keeping with the $SU(3)$ case as this is at the upper end of the 
conformal window. The lower end of the $SU(2)$ window is a current topic of
study which has not reached consensus yet, \cite{12,13}. For example, in 
\cite{12} the $\Nf$~$=$~$6$ value of $\rho$ is in the range $[0.26,0.74]$. Of
the schemes we have analysed only the three loop $\MOMq$ and $\MOMh$ estimates
lie comfortably within this band. This apparent agreement should be taken with 
caution due to the limit of perturbative credibility and lack of convergence as
well as the effect four loop corrections could have if the situation in
$\MSbar$ is a guide.  

Although the main interest in the Banks-Zaks fixed point stems from its
possible connection with a phase transition associated with chiral symmetry
breaking in QCD when the quarks are in the fundamental representation, for the
purposes of analysing possible theories beyond the Standard Model like 
\cite{17} we will consider the quarks in other representations. In this 
instance we will make brief remarks as there is mostly a parallel situation in 
these cases. When the quarks are in the adjoint representation the 
corresponding fixed point locations and critical exponents are provided in 
Tables $8$ to $13$. We stress that this not a supersymmetric version of QCD as 
there are not equal numbers of Bose and Fermi degrees of freedom. For the three
colour values we considered for the fundamental representation there is only 
one non-trivial infrared fixed point and then it is only present for
$\Nf$~$=$~$2$. In the MOM schemes for both $\omega$ and $\rho$ the same 
feature emerges in that the three loop estimates are $\Nc$ independent for the
$SU(\Nc)$ colour group. The $\Nc$ dependence becomes apparent at four loops
from Tables $10$ and $12$. For $\omega$ the three loop value of $\omega$ is 
around the same estimates of the four loop $\MSbar$ and $\mMOM$ values. By
contrast for $\rho$ the three loop $\MOMq$ estimates are competitive with the 
two four loop results except possibly for $\Nc$~$=$~$2$. This may be due to the
origin of the operator being a scalar quark bilinear. 

The next representation of interest is the $2S$ representation which 
corresponds to a double index symmetric representation. The results for this 
case are given in Tables $14$ to $19$ where there are only two fixed points for 
$\Nc$~$=$~$3$ and $4$ and again for low values of $\Nf$. The two critical 
exponents have similar properties to the fundamental representation case. For 
the larger of the two values of $\Nf$ there appears to be a convergent result 
when comparing the four loop results of \cite{17} and the three loop MOM 
scheme results except possibly for the $\MOMg$ scheme. For two flavours there 
is no clear pattern for either of the exponents or values of $\Nc$. Indeed for 
$\rho$ all bar one estimate is larger than unity. Finally, for the $2A$ 
representation, which is the antisymmetric double index partner to $2S$, the 
results are included in the Tables $20$ to $25$. While there are more fixed 
points for $\Nc$~$=$~$4$ we do not present results for $\Nc$~$=$~$3$. This is 
because in this representation the colour group Casimirs are precisely equal to
their corresponding values in the fundamental representation and we have 
commented on those results already. Though we do note that in the context of
model building or considering extensions to current theories quarks could be 
considered as being in the $2A$ representation rather than the fundamental one.
In terms of the critical exponents for $2A$ the situation for $\rho$ appears to
parallel our discussion for the differing behaviours of $\Nf$~$=$~$12$ and 
$13$. However, here the boundary appears to be at $\Nf$~$=$~$7$ and $8$. We 
would have to exclude the $\MOMg$ results from this analysis as it again seems 
to be an outlier for $\rho$. In terms of lattice analysis there has been an 
investigation for $\Nf$~$=$~$6$, \cite{11}, which is the lower boundary of the 
conformal window from the perturbative analysis. An estimate for $\rho$ lies in
the range $[0.3,0.35]$ for which only the four loop $\MSbar$ estimate is close 
to.  

We close this section by making some general comments on the analysis and try 
to give a perspective on the reliability of the perturbative estimates. In 
focusing the discussion so far on the comparison within a representation it may
miss some key features. For instance, for $\rho$ as a general rule it appears 
that when the value of $\rho_2$ is in the region of $1$ or larger then the 
higher loop estimates appear to be unreliable. By this we mean that the value 
appears to be at odds with estimates in other schemes. However, we need to be 
clear in saying this in that we are not suggesting that for that scheme the 
exponent does not converge. For values of $\Nf$ close to the upper boundary of 
the window in all the schemes the corresponding scheme estimates for $\rho$ 
clearly are in line with other schemes. What is probably the case is that more 
terms in the loop expansion for that particular scheme are needed in order to 
see the convergence. In the main the $\MOMg$ scheme appeared mostly to be in 
this outlier class. This is not unreasonable due to the nature of the $\MOMg$ 
scheme. It is based on ensuring that the triple gluon vertex has no $O(a)$ 
corrections at the completely symmetric point. Therefore, with the associated 
renormalization group functions their content is necessarily weighted by
gluonic rather than quark contributions. For the quark mass anomalous 
dimension, therefore, the quark content is not dominant. 

\sect{Discussion.}

It is worth making several general comments on our analysis. In \cite{17} the
evaluation of the quark mass anomalous dimensions at the Banks-Zaks fixed point
was examined in the conformal window for a set of renormalization schemes. We
have extended that analysis here to a different set of schemes which are the
momentum subtraction schemes of Celmaster and Gonsalves, \cite{43,44}. This is 
an important exercise since the analytic structure of the respective set of 
schemes is different from the point of view of the specific numbers which 
appear. Ultimately critical exponents which have been determined from the
renormalization group functions at criticality are renormalization group
invariants and the values have to be independent of the renormalization scheme
used to determine the anomalous dimensions. In this respect we have 
demonstrated this for the MOM QCD renormalization group functions at the
Wilson-Fisher fixed point in $d$~$=$~$4$~$-$~$2\epsilon$ dimensions. This is 
not a trivial exercise as the $d$-dimensional renormalization group functions 
are required in the MOM case in order to observe the renormalization group 
invariance in $d$-dimensions. For the Banks-Zaks fixed point the situation is 
different with regard to the invariance. Until the quantum field theory which 
drives the Banks-Zaks fixed point is found then at present a numerical 
evaluation of the critical exponents order by order in the loop expansion is 
the only tool available. In other words there will be a theory in the same 
universality class as QCD at the Banks-Zaks fixed point where direct 
computation of its anomalous dimensions in various schemes ought to be the way 
to see the renormalization invariance of the critical exponents. Having said 
this on the whole, despite the differing numeric natures of the renormalization
group functions in MOM schemes versus those of the $\MSbar$, RI${}^\prime$ and 
$\mMOM$ schemes analysed in \cite{17}, the scheme dependence appears to 
disappear for values of $\Nf$ near the upper end of the conformal window for
the various quark representations we have considered. This is where
perturbation theory is at its most reliable. One interesting point is when
there are $\Nf$~$=$~$12$ fundamental flavours for $SU(3)$. On the whole the 
quark mass anomalous dimension appears to be converging slowly towards recent 
values measured on the lattice, \cite{14,15}. For the $\MOMq$ scheme the 
{\em three} loop estimate of $\rho$ is closer than the corresponding $\MSbar$ 
value. Whether there is faster convergence for this particular scheme remains 
to be seen in the absence of a full four loop computation. Given the nature of 
this scheme, which is founded on the quark-gluon vertex, it may be the case 
that the quark mass anomalous dimension in this scheme does indeed have the 
best convergence. However, these remarks need to be tempered by the 
observations in \cite{17} where it was noted that $\Nf$~$=$~$12$ may be the 
point where non-perturbative features become dominant. A measure of that can be
seen in the evaluation of the stability critical exponent $\omega$. In Tables 
$3$ and $4$ for $\Nf$~$=$~$13$ the value of $\omega$ appears to be consistent 
across all the schemes considered except for $\MOMg$. The values for $\rho$ for
the same $\Nf$ accord with this. For $\Nf$~$=$~$12$ the estimates of $\omega$ 
have a broader range across the schemes. 

\vspace{1cm}
\noindent
{\bf Acknowledgements.} This work was carried out with the support of STFC 
including an STFC studentship (RMS). We thank Dr T.A. Ryttov for discussions.

\clearpage
{\begin{table}[hb]
\begin{center}
\begin{tabular}{|c|c||r|r|r||r|r|r|}
\hline
& $F$ & & $\MSbars$ & & & $\mMOM$ & \\
\hline
$\Nc$ & $\Nf$ & $a_2$ & $a_3$ & $a_4$ & 
$a_2$ & $a_3$ & $a_4$ \\
\hline
2 & 6 & 0.909091 & 0.130937 & 0.190588 &
0.909091 & 0.100122 & 0.088677 \\
2 & 7 & 0.225352 & 0.083898 & 0.096318 &
0.225352 & 0.067933 & 0.062904 \\
2 & 8 & 0.100000 & 0.054773 & 0.060487 &
0.100000 & 0.046821 & 0.045404 \\
2 & 9 & 0.047337 & 0.033280 & 0.035339 &
0.047337 & 0.030031 & 0.029984 \\
2 & 10 & 0.018349 & 0.015622 & 0.015944 & 
0.018349 & 0.014878 & 0.014954 \\
\hline
3 & 9 & 0.416667 & 0.081803 & 0.085291 &
0.416667 & 0.064438 & 0.054935 \\
3 & 10 & 0.175676 & 0.060824 & 0.064860 & 
0.175676 & 0.049421 & 0.044230 \\
3 & 11 & 0.098214 & 0.046039 & 0.049832 &
0.098214 & 0.038603 & 0.036070 \\
3 & 12 & 0.060000 & 0.034607 & 0.037434 &
0.060000 & 0.029962 & 0.028981 \\
3 & 13 & 0.037234 & 0.025191 & 0.026853 & 
0.037234 & 0.022535 & 0.022329 \\
3 & 14 & 0.022124 & 0.017070 & 0.017793 &
0.022124 & 0.015786 & 0.015838 \\
3 & 15 & 0.011364 & 0.009818 & 0.010001 &
0.011364 & 0.009383 & 0.009431 \\
3 & 16 & 0.003311 & 0.003162 & 0.003170 & 
0.003311 & 0.003118 & 0.003121 \\
\hline
4 & 12 & 0.281690 & 0.060040 & 0.060411 &
0.281690 & 0.047748 & 0.040336 \\
4 & 13 & 0.147239 & 0.048027 & 0.049944 &
0.147239 & 0.039016 & 0.034347 \\
4 & 14 & 0.092219 & 0.038926 & 0.041445 &
0.092219 & 0.032328 & 0.029529 \\
4 & 15 & 0.062291 & 0.031616 & 0.034072 &
0.062291 & 0.026858 & 0.025323 \\
4 & 16 & 0.043478 & 0.025488 & 0.027490 &
0.043478 & 0.022159 & 0.021442 \\
4 & 17 & 0.030558 & 0.020179 & 0.021580 &
0.030558 & 0.017964 & 0.017724 \\
4 & 18 & 0.021136 & 0.015460 & 0.016291 &
0.021136 & 0.014097 & 0.014086 \\
4 & 19 & 0.013962 & 0.011175 & 0.011573 &
0.013962 & 0.010440 & 0.010493 \\
4 & 20 & 0.008316 & 0.007218 & 0.007350 &
0.008316 & 0.006907 & 0.006943 \\
4 & 21 & 0.003758 & 0.003511 & 0.003530 &
0.003758 & 0.003438 & 0.003446 \\
\hline
\end{tabular}
\end{center}
%\vspace{0.3cm}
\begin{center}
{Table $1$. Location of Banks-Zaks critical points for $\MSbar$ and $\mMOM$ at 
two, three and four loops.}
\end{center}
\end{table}}

\clearpage
{\begin{table}[ht]
\begin{center}
\begin{tabular}{|c|c||r|r||r|r||r|r|}
\hline
& $F$ & & $\MOMq$ & & $\MOMg$ & & $\MOMh$ \\
\hline
$\Nc$ & $\Nf$ & $a_2$ & $a_3$ & $a_2$ & $a_3$ & $a_2$ & $a_3$ \\
\hline
2 & 6 & 0.909091 & 0.079453 &
0.909091 & 0.075345 &
0.909091 & 0.100010 \\
2 & 7 & 0.225352 & 0.060047 &
0.225352 & 0.051522 &
0.225352 & 0.069384 \\
2 & 8 & 0.100000 & 0.044163 &
0.100000 & 0.035988 &
0.100000 & 0.048379 \\
2 & 9 & 0.047337 & 0.029574 &
0.047337 & 0.023848 &
0.047337 & 0.031152 \\
2 & 10 & 0.018349 & 0.014999 &
0.018349 & 0.012674 &
0.018349 & 0.015317 \\
\hline
3 & 9 & 0.416667 & 0.051906 &
0.416667 & 0.047997 &
0.416667 & 0.064858 \\
3 & 10 & 0.175676 & 0.042853 &
0.175676 & 0.037161 &
0.175676 & 0.050466 \\
3 & 11 & 0.098214 & 0.035202 &
0.098214 & 0.029277 &
0.098214 & 0.039778 \\
3 & 12 & 0.060000 & 0.028357 &
0.060000 & 0.023018 &
0.060000 & 0.031047 \\
3 & 13 & 0.037234 & 0.021938 &
0.037234 & 0.017681 &
0.037234 & 0.023405 \\
3 & 14 & 0.022124 & 0.015687 &
0.022124 & 0.012809 &
0.022124 & 0.016367 \\
3 & 15 & 0.011364 & 0.009437 &
0.011364 & 0.008032 &
0.011364 & 0.009655 \\
3 & 16 & 0.003311 & 0.003136 &
0.003311 & 0.002914 &
0.003311 & 0.003156 \\
\hline
4 & 12 & 0.281690 & 0.038650 &
0.281690 & 0.035451 &
0.281690 & 0.048181 \\
4 & 13 & 0.147239 & 0.033425 &
0.147239 & 0.029214 &
0.147239 & 0.039802 \\
4 & 14 & 0.092219 & 0.028879 &
0.092219 & 0.024372 &
0.092219 & 0.033229 \\
4 & 15 & 0.062291 & 0.024786 &
0.062291 & 0.020409 &
0.062291 & 0.027756 \\
4 & 16 & 0.043478 & 0.020992 &
0.043478 & 0.017023 &
0.043478 & 0.022982 \\
4 & 17 & 0.030558 & 0.017383 &
0.030558 & 0.014013 &
0.030558 & 0.018663 \\
4 & 18 & 0.021136 & 0.013876 &
0.021136 & 0.011238 &
0.021136 & 0.014641 \\
4 & 19 & 0.013962 & 0.010408 &
0.013962 & 0.008578 &
0.013962 & 0.010810 \\
4 & 20 & 0.008316 & 0.006940 &
0.008316 & 0.005922 &
0.008316 & 0.007105 \\
4 & 21 & 0.003758 & 0.003461 &
0.003758 & 0.003134 &
0.003758 & 0.003498 \\
\hline
\end{tabular}
\end{center}
%\vspace{0.3cm}
\begin{center}
{Table $2$. Location of Banks-Zaks critical points for $\MOMq$, $\MOMg$ and 
$\MOMh$ at two and three loops.}
\end{center}
\end{table}}

\clearpage
{\begin{table}[ht]
\begin{center}
\begin{tabular}{|c|c||r|r|r||r|r|r|}
\hline
& $F$ & & $\MSbars$ & & & $\mMOM$ & \\
\hline
$\Nc$ & $\Nf$ & $\omega_2$ & $\omega_3$ & $\omega_4$ & 
$\omega_2$ & $\omega_3$ & $\omega_4$ \\
\hline
2 & 6 & 6.060606 & 1.620106 & 0.974775 &
6.060606 & 1.261453 & 1.245537 \\
2 & 7 & 1.201878 & 0.728326 & 0.676986 &
1.201878 & 0.615403 & 0.618233 \\
2 & 8 & 0.400000 & 0.318182 & 0.299703 &
0.400000 & 0.286878 & 0.289100 \\
2 & 9 & 0.126233 & 0.115100 & 0.110454 &
0.126233 & 0.109360 & 0.109439 \\
2 & 10 & 0.024465 & 0.023925 & 0.023541 &
0.024465 & 0.023590 & 0.023507 \\
\hline
3 & 9 & 4.166667 & 1.475455 & 1.464386 &
4.166667 & 1.189101 & 1.165667 \\
3 & 10 & 1.522523 & 0.871775 & 0.853407 &
1.522533 & 0.736141 & 0.736306 \\
3 & 11 & 0.720238 & 0.516977 & 0.498035 &
0.720238 & 0.454913 & 0.459085 \\
3 & 12 & 0.360000 & 0.295517 & 0.282328 &
0.360000 & 0.269774 & 0.272234 \\
3 & 13 & 0.173759 & 0.155581 & 0.149130 &
0.173759 & 0.146681 & 0.147243 \\
3 & 14 & 0.073746 & 0.069899 & 0.067812 &
0.073746 & 0.067695 & 0.067572 \\
3 & 15 & 0.022727 & 0.022307 & 0.021975 &
0.022727 & 0.022037 & 0.021957 \\
3 & 16 & 0.002208 & 0.002203 & 0.002198 &
0.002208 & 0.002200 & 0.002198 \\
\hline
4 & 12 & 3.755869 & 1.430447 & 1.429308 &
3.755897 & 1.165365 & 1.140669 \\
4 & 13 & 1.766871 & 0.964661 & 0.954675 &
1.766861 & 0.812318 & 0.809419 \\
4 & 14 & 0.983670 & 0.655163 & 0.639277 &
0.983670 & 0.568776 & 0.572539 \\
4 & 15 & 0.581387 & 0.440398 & 0.424261 &
0.581387 & 0.393264 & 0.397364 \\
4 & 16 & 0.347826 & 0.288274 & 0.275809 &
0.347826 & 0.264197 & 0.266663 \\
4 & 17 & 0.203718 & 0.180219 & 0.172523 &
0.203718 & 0.169115 & 0.170002 \\
4 & 18 & 0.112726 & 0.104596 & 0.100807 &
0.112726 & 0.100224 & 0.100263 \\
4 & 19 & 0.055846 & 0.053622 & 0.052223 &
0.055846 & 0.052293 & 0.052131 \\
4 & 20 & 0.022176 & 0.021789 & 0.021468 &
0.022176 & 0.021539 & 0.021457 \\
4 & 21 & 0.005010 & 0.004989 & 0.004965 &
0.005010 & 0.004974 & 0.004964 \\
\hline
\end{tabular}
\end{center}
%\vspace{0.3cm}
\begin{center}
{Table $3$. Critical exponent $\omega$ for the Banks-Zaks critical point for
$\MSbar$ and $\mMOM$ at two, three and four loops.}
\end{center}
\end{table}}

\clearpage
{\begin{table}[ht]
\begin{center}
\begin{tabular}{|c|c||r|r||r|r||r|r|}
\hline
& $F$ & & $\MOMq$ & & $\MOMg$ & & $\MOMh$ \\
\hline
$\Nc$ & $\Nf$ & $\omega_2$ & $\omega_3$ & $\omega_2$ & $\omega_3$ & 
$\omega_2$ & $\omega_3$ \\
\hline
2 & 6 & 6.060606 & 1.013077 &
6.060606 & 0.962970 &
6.060606 & 1.260113 \\
2 & 7 & 1.201878 & 0.555171 &
1.201878 & 0.486742 &
1.201878 & 0.626165 \\
2 & 8 & 0.400000 & 0.275290 &
0.400000 & 0.236097 &
0.400000 & 0.293412 \\
2 & 9 & 0.126233 & 0.108457 &
0.126233 & 0.095150 &
0.126233 & 0.111475 \\
2 & 10 & 0.024465 & 0.023649 &
0.024465 & 0.022125 &
0.024465 & 0.023797 \\
\hline
3 & 9 & 4.166667 & 0.973459 &
4.166667 & 0.904648 &
4.166667 & 1.196201 \\
3 & 10 & 1.522523 & 0.652189 &
1.522533 & 0.575996 &
1.522533 & 0.749100 \\
3 & 11 & 0.720238 & 0.423769 &
0.720238 & 0.365393 &
0.720238 & 0.465266 \\
3 & 12 & 0.360000 & 0.259872 &
0.360000 & 0.223235 &
0.360000 & 0.276171 \\
3 & 13 & 0.173759 & 0.144437 &
0.173759 & 0.125839 &
0.173759 & 0.149791 \\
3 & 14 & 0.073746 & 0.067504 &
0.073746 & 0.060674 &
0.073746 & 0.068753 \\
3 & 15 & 0.022727 & 0.022074 &
0.022727 & 0.020774 &
0.022727 & 0.022213 \\
3 & 16 & 0.002208 & 0.002201 &
0.002208 & 0.002176 &
0.002208 & 0.002203 \\
\hline
4 & 12 & 3.755869 & 0.959967 &
3.755869 & 0.885870 &
3.755869 & 1.174951 \\
4 & 13 & 1.766871 & 0.711138 &
1.766871 & 0.631571 &
1.766871 & 0.826128 \\
4 & 14 & 0.983670 & 0.519614 &
0.983670 & 0.451225 &
0.983670 & 0.581177 \\
4 & 15 & 0.581387 & 0.370624 &
0.581387 & 0.318564 &
0.581387 & 0.402677 \\
4 & 16 & 0.347826 & 0.254787 &
0.347826 & 0.219046 &
0.347826 & 0.270526 \\
4 & 17 & 0.203718 & 0.165850 &
0.203718 & 0.144003 &
0.203718 & 0.172852 \\
4 & 18 & 0.112726 & 0.099425 &
0.112726 & 0.088003 &
0.112726 & 0.102081 \\
4 & 19 & 0.055846 & 0.052229 &
0.055846 & 0.047543 &
0.055846 & 0.053001 \\
4 & 20 & 0.022176 & 0.021569 &
0.022176 & 0.020338 &
0.022176 & 0.021706 \\
4 & 21 & 0.005010 & 0.004979 &
0.005010 & 0.004872 &
0.005010 & 0.004986 \\
\hline
\end{tabular}
\end{center}
%\vspace{0.3cm}
\begin{center}
{Table $4$. Critical exponent $\omega$ for Banks-Zaks critical point for 
$\MOMq$, $\MOMg$ and $\MOMh$ at two and three loops.}
\end{center}
\end{table}}

\clearpage
{\begin{table}[ht]
\begin{center}
\begin{tabular}{|c|c||r|r|r||r|r|r||r|}
\hline
& $F$ & & $\MSbars$ & & & $\mMOM$ & & $\mbox{'t Hooft}$ \\
\hline
$\Nc$ & $\Nf$ & $\rho_2$ & $\rho_3$ & $\rho_4$ & 
$\rho_2$ & $\rho_3$ & $\rho_4$ & $\rho$ \\
\hline
2 & 6 & 33.171488 & 0.924853 & - 4.019013 &
39.576446 & 1.034933 & 0.893430 & 4.090909 \\
2 & 7 & 2.674073 & 0.456824 & 0.032536 &
3.118429 & 0.523238 & 0.455155 & 1.014085 \\
2 & 8 & 0.751875 & 0.272074 & 0.203618 &
0.849375 & 0.300337 & 0.279549 & 0.450000 \\
2 & 9 & 0.275060 & 0.160546 & 0.157402 &
0.299149 & 0.168800 & 0.165956 & 0.213018 \\
2 & 10 & 0.091049 & 0.073829 & 0.074794 &
0.095005 & 0.074836 & 0.075064 & 0.082569 \\
\hline
3 & 9 & 19.768519 & 1.061659 & - 0.143490 &
23.356481 & 1.191042 & 0.979184 & 3.333333 \\
3 & 10 & 4.189838 & 0.646806 & 0.155885 &
4.882518 & 0.734781 & 0.620806 & 1.405405 \\
3 & 11 & 1.613131 & 0.439241 & 0.249686 &
1.846779 & 0.492300 & 0.436592 & 0.785714 \\
3 & 12 & 0.772800 & 0.311751 & 0.253328 &
0.866400 & 0.340313 & 0.317156 & 0.480000 \\
3 & 13 & 0.404469 & 0.220154 & 0.209757 &
0.442979 & 0.233293 & 0.226367 & 0.297872 \\
3 & 14 & 0.212450 & 0.146369 & 0.147421 &
0.226917 & 0.151029 & 0.150241 & 0.176991 \\
3 & 15 & 0.099690 & 0.082573 & 0.083600 &
0.103736 & 0.083547 & 0.083816 & 0.090909 \\
3 & 16 & 0.027187 & 0.025833 & 0.025895 &
0.027550 & 0.025868 & 0.025896 & 0.026490 \\
\hline
4 & 12 & 17.296915 & 1.107600 & 0.058357 &
20.371702 & 1.243981 & 1.009616 & 3.169014 \\
4 & 13 & 5.380895 & 0.755292 & 0.192015 &
6.275170 & 0.855872 & 0.712621 & 1.656442 \\
4 & 14 & 2.445332 & 0.552297 & 0.258813 &
2.817397 & 0.622351 & 0.537602 & 1.037464 \\
4 & 15 & 1.318886 & 0.420081 & 0.280672 &
1.498346 & 0.466289 & 0.419073 & 0.700779 \\
4 & 16 & 0.778444 & 0.324942 & 0.268806 &
0.870599 & 0.353508 & 0.329838 & 0.489130 \\
4 & 17 & 0.480849 & 0.250606 & 0.234022 &
0.528704 & 0.266804 & 0.256937 & 0.343774 \\
4 & 18 & 0.300568 & 0.188596 & 0.186947 &
0.324580 & 0.196704 & 0.193870 & 0.237781 \\
4 & 19 & 0.183246 & 0.134334 & 0.136002 &
0.194211 & 0.137668 & 0.137526 & 0.157068 \\
4 & 20 & 0.102410 & 0.085397 & 0.086461 &
0.106473 & 0.086356 & 0.086657 & 0.093555 \\
4 & 21 & 0.043993 & 0.040685 & 0.040877 &
0.044858 & 0.040801 & 0.040884 & 0.042273 \\
\hline
\end{tabular}
\end{center}
%\vspace{0.3cm}
\begin{center}
{Table $5$. Quark mass critical exponent at the Banks-Zaks critical point for 
the $\MSbar$ and $\mMOM$ schemes at two, three and four loops and the 't Hooft
scheme.}
\end{center}
\end{table}}

\clearpage
{\begin{table}
\begin{center}
\begin{tabular}{|c|c||r|r||r|r||r|r|}
\hline
& $F$ & & $\MOMq$ & & $\MOMg$ & & $\MOMh$ \\
\hline
$\Nc$ & $\Nf$ & $\rho_2$ & $\rho_3$ & $\rho_2$ & $\rho_3$ & 
$\rho_2$ & $\rho_3$ \\
\hline
2 & 6 & 17.262397 & 0.461381 &
45.730994 & 0.861480 &
13.977991 & 0.305679 \\
2 & 7 & 1.925820 & 0.346755 &
4.202074 & 0.542352 &
1.724000 & 0.304515 \\
2 & 8 & 0.649692 & 0.247039 &
1.201675 & 0.336642 &
0.609951 & 0.234852 \\
2 & 9 & 0.262282 & 0.156674 &
0.409221 & 0.189326 &
0.253377 & 0.153796 \\
2 & 10 & 0.090649 & 0.073686 &
0.116219 & 0.079602 &
0.089311 & 0.073373 \\
\hline
3 & 9 & 11.561746 & 0.553462 &
26.804208 & 0.954324 &
9.016606 & 0.375534 \\
3 & 10 & 2.978729 & 0.452897 &
6.257561 & 0.701362 &
2.526293 & 0.377682 \\
3 & 11 & 1.312033 & 0.364656 &
2.514774 & 0.516777 &
1.170622 & 0.330763 \\
3 & 12 & 0.689329 & 0.285218 &
1.204608 & 0.374024 &
0.636553 & 0.270097 \\
3 & 13 & 0.383454 & 0.212345 &
0.607462 & 0.259356 &
0.363130 & 0.206167 \\
3 & 14 & 0.208960 & 0.144860 &
0.297076 & 0.165375 &
0.201785 & 0.142818 \\
3 & 15 & 0.099806 & 0.082504 &
0.125435 & 0.088262 &
0.097913 & 0.082094 \\
3 & 16 & 0.027285 & 0.025840 &
0.029663 & 0.026154 &
0.027124 & 0.025826 \\
\hline
4 & 12 & 10.475472 & 0.586353 &
23.326276 & 0.984386 &
8.082819 & 0.401265 \\
4 & 13 & 3.761930 & 0.503058 &
7.835301 & 0.780943 &
3.108222 & 0.406260 \\
4 & 14 & 1.906259 & 0.428513 &
3.724747 & 0.622031 &
1.649824 & 0.375454 \\
4 & 15 & 1.116733 & 0.360679 &
2.047092 & 0.493265 &
0.999731 & 0.331200 \\
4 & 16 & 0.701301 & 0.298087 &
1.203586 & 0.386035 &
0.644300 & 0.281973 \\
4 & 17 & 0.453285 & 0.239729 &
0.725617 & 0.294985 &
0.425128 & 0.231347 \\
4 & 18 & 0.292424 & 0.184975 &
0.434301 & 0.216727 &
0.278953 & 0.181016 \\
4 & 19 & 0.181893 & 0.133522 &
0.248856 & 0.149155 &
0.176016 & 0.131953 \\
4 & 20 & 0.102711 & 0.085353 &
0.128262 & 0.091032 &
0.100626 & 0.084913 \\
4 & 21 & 0.044214 & 0.040704 &
0.049797 & 0.041635 &
0.043788 & 0.040652 \\
\hline
\end{tabular}
\end{center}
%\vspace{0.3cm}
\begin{center}
{Table $6$. Quark mass critical exponent at the Banks-Zaks critical point for 
$\MOMq$, $\MOMg$ and $\MOMh$ at two and three loops.}
\end{center}
\end{table}}

{\begin{table}
\begin{center}
\begin{tabular}{|c|c||r|r|}
\hline
& $F$ & & $\MSbars$ \\
\hline
$\Nc$ & $\Nf$ & $\rho_{53}$ & $\rho_{54}$ \\ 
\hline
3 & 9 & - 0.370415 & - 0.596381 \\
3 & 10 & 0.198718 & 0.105449 \\
3 & 11 & 0.289590 & 0.266959 \\
3 & 12 & 0.262582 & 0.268132 \\
3 & 13 & 0.205572 & 0.215243 \\
3 & 14 & 0.143001 & 0.148548 \\
3 & 15 & 0.082153 & 0.083692 \\
3 & 16 & 0.025828 & 0.025895 \\
\hline
\end{tabular}
\end{center}
%\vspace{0.3cm}
\begin{center}
{Table $7$. Estimates of quark mass critical exponent at the Banks-Zaks 
critical point for the $\MSbar$ at five loops using the three and four loop
critical coupling.}
\end{center}
\end{table}}

\clearpage
{\begin{table}
\begin{center}
\begin{tabular}{|c|c||r|r|r||r|r|r|}
\hline
& $G$ & & $\MSbars$ & & & $\mMOM$ & \\
\hline
$\Nc$ & $\Nf$ & $a_2$ & $a_3$ & $a_4$ & 
$a_2$ & $a_3$ & $a_4$ \\
\hline
2 & 2 & 0.050000 & 0.036525 & 0.035814 & 
0.050000 & 0.033778 & 0.031703 \\
3 & 2 & 0.033333 & 0.024350 & 0.024537 &
0.033333 & 0.022519 & 0.021491 \\
4 & 2 & 0.025000 & 0.018263 & 0.018596 &
0.025000 & 0.016889 & 0.016217 \\
\hline
\end{tabular}
\end{center}
%\vspace{0.3cm}
\begin{center}
{Table $8$. Location of Banks-Zaks critical points for $\MSbar$ and $\mMOM$ at 
two, three and four loops for the quarks in the adjoint representation.}
\end{center}
\end{table}}

{\begin{table}
\begin{center}
\begin{tabular}{|c|c||r|r||r|r||r|r|}
\hline
& $G$ & & $\MOMq$ & & $\MOMg$ & & $\MOMh$ \\
\hline
$\Nc$ & $\Nf$ & $a_2$ & $a_3$ & $a_2$ & $a_3$ & $a_2$ & $a_3$ \\
\hline
2 & 2 & 0.050000 & 0.032037 
& 0.050000 & 0.026198 & 0.050000 & 0.035416 \\
3 & 2 & 0.033333 & 0.021358 
& 0.033333 & 0.017465 & 0.033333 & 0.023611 \\
4 & 2 & 0.025000 & 0.016019 
& 0.025000 & 0.013099 & 0.025000 & 0.017708 \\
\hline
\end{tabular}
\end{center}
%\vspace{0.3cm}
\begin{center}
{Table $9$. Location of Banks-Zaks critical points for $\MOMq$, $\MOMg$ and 
$\MOMh$ at two and three loops for the quarks in the adjoint representation.}
\end{center}
\end{table}}

{\begin{table}
\begin{center}
\begin{tabular}{|c|c||r|r|r||r|r|r|}
\hline
& $G$ & & $\MSbars$ & & & $\mMOM$ & \\
\hline
$\Nc$ & $\Nf$ & $\omega_2$ & $\omega_3$ & $\omega_4$ & 
$\omega_2$ & $\omega_3$ & $\omega_4$ \\
\hline
2 & 2 & 0.200000 & 0.185475 & 0.187427 &
0.200000 & 0.178949 & 0.183383 \\
3 & 2 & 0.200000 & 0.185475 & 0.184637 & 
0.200000 & 0.178949 & 0.182466 \\
4 & 2 & 0.200000 & 0.185475 & 0.183419 &
0.200000 & 0.178949 & 0.182086 \\
\hline
\end{tabular}
\end{center}
%\vspace{0.3cm}
\begin{center}
{Table $10$. Critical exponent $\omega$ for the Banks-Zaks critical point for
$\MSbar$ and $\mMOM$ at two, three and four loops for the quarks in the adjoint
representation.}
\end{center}
\end{table}}

{\begin{table}
\begin{center}
\begin{tabular}{|c|c||r|r||r|r||r|r|}
\hline
& $G$ & & $\MOMq$ & & $\MOMg$ & & $\MOMh$ \\
\hline
$\Nc$ & $\Nf$ & $\omega_2$ & $\omega_3$ & $\omega_2$ & $\omega_3$ & 
$\omega_2$ & $\omega_3$ \\
\hline
2 & 2 & 0.200000 & 0.174187 & 0.200000 & 0.154678 &
0.200000 & 0.182985 \\
3 & 2 & 0.200000 & 0.174187 & 0.200000 & 0.154678 &
0.200000 & 0.182985 \\ 
4 & 2 & 0.200000 & 0.174187 & 0.200000 & 0.154678 &
0.200000 & 0.182985 \\
\hline
\end{tabular}
\end{center}
%\vspace{0.3cm}
\begin{center}
{Table $11$. Critical exponent $\omega$ for Banks-Zaks critical point for 
$\MOMq$, $\MOMg$ and $\MOMh$ at two and three loops for the quarks in the
adjoint representation.}
\end{center}
\end{table}}

\clearpage
{\begin{table}
\begin{center}
\begin{tabular}{|c|c||r|r|r||r|r|r|}
\hline
& $G$ & & $\MSbars$ & & & $\mMOM$ & \\
\hline
$\Nc$ & $\Nf$ & $\rho_2$ & $\rho_3$ & $\rho_4$ & 
$\rho_2$ & $\rho_3$ & $\rho_4$ \\
\hline
2 & 2 & 0.820000 & 0.543233 & 0.499621 &
0.885000 & 0.569034 & 0.520679 \\
3 & 2 & 0.820000 & 0.543233 & 0.522652 &
0.885000 & 0.569034 & 0.537795 \\
4 & 2 & 0.820000 & 0.543233 & 0.531736 &
0.885000 & 0.569034 & 0.544255 \\
\hline
\end{tabular}
\end{center}
%\vspace{0.3cm}
\begin{center}
{Table $12$. Quark mass critical exponent at the Banks-Zaks critical point for 
the $\MSbar$ and $\mMOM$ schemes at two, three and four loops for the quarks in
the adjoint representation.}
\end{center}
\end{table}}

{\begin{table}
\begin{center}
\begin{tabular}{|c|c||r|r||r|r||r|r|}
\hline
& $G$ & & $\MOMq$ & & $\MOMg$ & & $\MOMh$ \\
\hline
$\Nc$ & $\Nf$ & $\rho_2$ & $\rho_3$ & $\rho_2$ & $\rho_3$ & 
$\rho_2$ & $\rho_3$ \\
\hline
2 & 2 & 0.843280 & 0.523076 & 1.119867 & 0.563241 &
0.725384 & 0.493780 \\
3 & 2 & 0.843279 & 0.523076 & 1.119867 & 0.563241 &
0.725384 & 0.493780 \\ 
4 & 2 & 0.843280 & 0.523076 & 1.119867 & 0.563241 &
0.725384 & 0.493780 \\
\hline
\end{tabular}
\end{center}
%\vspace{0.3cm}
\begin{center}
{Table $13$. Quark mass critical exponent at the Banks-Zaks critical point for 
$\MOMq$, $\MOMg$ and $\MOMh$ at two and three loops for the quarks in the 
adjoint representation.}
\end{center}
\end{table}}

{\begin{table}
\begin{center}
\begin{tabular}{|c|c||r|r|r||r|r|r|}
\hline
& $2S$ & & $\MSbars$ & & & $\mMOM$ & \\
\hline
$\Nc$ & $\Nf$ & $a_2$ & $a_3$ & $a_4$ & 
$a_2$ & $a_3$ & $a_4$ \\
\hline
3 & 2 & 0.067010 & 0.039795 & 0.037400 &
0.067010 & 0.036641 & 0.031345 \\
3 & 3 & 0.006757 & 0.006290 & 0.006324 & 
0.006757 & 0.006133 & 0.006137 \\
\hline
4 & 2 & 0.076923 & 0.038610 & 0.034993 &
0.076923 & 0.035879 & 0.028481 \\
4 & 3 & 0.012085 & 0.010266 & 0.010429 & 
0.012085 & 0.009773 & 0.009706 \\
\hline
\end{tabular}
\end{center}
%\vspace{0.3cm}
\begin{center}
{Table $14$. Location of Banks-Zaks critical points for $\MSbar$ and $\mMOM$ at 
two, three and four loops for quarks in the $2S$ representation.}
\end{center}
\end{table}}

{\begin{table}
\begin{center}
\begin{tabular}{|c|c||r|r||r|r||r|r|}
\hline
& $2S$ & & $\MOMq$ & & $\MOMg$ & & $\MOMh$ \\
\hline
$\Nc$ & $\Nf$ & $a_2$ & $a_3$ & $a_2$ & 
$a_3$ & $a_2$ & $a_3$ \\
\hline
3 & 2 & 0.067010 & 0.033185 & 0.067010 &
0.026936 & 0.067010 & 0.038706 \\
3 & 3 & 0.006757 & 0.006043 & 0.006757 &
0.005449 & 0.006757 & 0.006272 \\
\hline
4 & 2 & 0.076923 & 0.031380 & 0.076923 &
0.025513 & 0.076923 & 0.037977 \\
4 & 3 & 0.012085 & 0.009452 & 0.012085 &
0.008038 & 0.012085 & 0.010154 \\
\hline
\end{tabular}
\end{center}
%\vspace{0.3cm}
\begin{center}
{Table $15$. Location of Banks-Zaks critical points for $\MOMq$, $\MOMg$ and 
$\MOMh$ at two and three loops for quarks in the $2S$ representation.}
\end{center}
\end{table}}

\clearpage 
{\begin{table}
\begin{center}
\begin{tabular}{|c|c||r|r|r||r|r|r|}
\hline
& $2S$ & & $\MSbars$ & & & $\mMOM$ & \\
\hline
$\Nc$ & $\Nf$ & $\omega_2$ & $\omega_3$ & $\omega_4$ & 
$\omega_2$ & $\omega_3$ & $\omega_4$ \\
\hline
3 & 2 & 0.580756 & 0.484962 & 0.494313 &
0.580756 & 0.461475 & 0.470733 \\
3 & 3 & 0.013514 & 0.013449 & 0.013385 &
0.013514 & 0.013398 & 0.013391 \\
\hline
4 & 2 & 1.025641 & 0.771209 & 0.784341 &
1.025641 & 0.733643 & 0.730358 \\
4 & 3 & 0.064451 & 0.062991 & 0.062225 &
0.064451 & 0.062094 & 0.062379 \\
\hline
\end{tabular}
\end{center}
%\vspace{0.3cm}
\begin{center}
{Table $16$. Critical exponent $\omega$ for the Banks-Zaks critical point for 
$\MSbar$ and $\mMOM$ at two, three and four loops for quarks in the $2S$ 
representation.}
\end{center}
\end{table}}

{\begin{table}
\begin{center}
\begin{tabular}{|c|c||r|r||r|r||r|r|}
\hline
& $2S$ & & $\MOMq$ & & $\MOMg$ & & $\MOMh$ \\
\hline
$\Nc$ & $\Nf$ & $\omega_2$ & $\omega_3$ & $\omega_2$ & 
$\omega_3$ & $\omega_2$ & $\omega_3$ \\
\hline
3 & 2 & 0.580756 & 0.432782 & 0.580756 &
0.373054 & 0.580756 & 0.477139 \\
3 & 3 & 0.013514 & 0.013363 & 0.013514 &
0.013007 & 0.013514 & 0.013444 \\
\hline
4 & 2 & 1.025641 & 0.666122 & 1.025641 &
0.567521 & 1.025641 & 0.762733 \\
4 & 3 & 0.064451 & 0.061393 & 0.064451 &
0.057224 & 0.064451 & 0.062807 \\
\hline
\end{tabular}
\end{center}
%\vspace{0.3cm}
\begin{center}
{Table $17$. Critical exponent $\omega$ for the Banks-Zaks critical point for 
the $\MOMq$, $\MOMg$ and $\MOMh$ schemes at two and three loops for quarks in 
the $2S$ representation.}
\end{center}
\end{table}}

{\begin{table}
\begin{center}
\begin{tabular}{|c|c||r|r|r||r|r|r|}
\hline
& $2S$ & & $\MSbars$ & & & $\mMOM$ & \\
\hline
$\Nc$ & $\Nf$ & $\rho_2$ & $\rho_3$ & $\rho_4$ & 
$\rho_2$ & $\rho_3$ & $\rho_4$ \\
\hline
3 & 2 & 2.442844 & 1.284021 & 1.122151 &
2.694805 & 1.422422 & 1.210883 \\
3 & 3 & 0.143809 & 0.132625 & 0.133158 &
0.147386 & 0.133175 & 0.133159 \\
\hline
4 & 2 & 4.815089 & 2.077658 & 1.787181 &
5.365385 & 2.436574 & 1.949337 \\
4 & 3 & 0.380719 & 0.313071 & 0.314964 &
0.399558 & 0.318680 & 0.315594 \\
\hline
\end{tabular}
\end{center}
%\vspace{0.3cm}
\begin{center}
{Table $18$. Quark mass critical exponent at the Banks-Zaks critical point for 
the $\MSbar$ and $\mMOM$ schemes at two, three and four loops for quarks in the
$2S$ representation.}
\end{center}
\end{table}}

{\begin{table}
\begin{center}
\begin{tabular}{|c|c||r|r||r|r||r|r|}
\hline
& $2S$ & & $\MOMq$ & & $\MOMg$ & & $\MOMh$ \\
\hline
$\Nc$ & $\Nf$ & $\rho_2$ & $\rho_3$ & $\rho_2$ & 
$\rho_3$ & $\rho_2$ & $\rho_3$ \\
\hline
3 & 2 & 2.440100 & 1.088873 & 3.194973 &
1.123601 & 1.837734 & 0.959833 \\
3 & 3 & 0.148363 & 0.133049 & 0.166564 &
0.135940 & 0.142239 & 0.132247 \\
\hline
4 & 2 & 4.616444 & 1.554419 & 5.894166 &
1.548531 & 3.166038 & 1.294776 \\
4 & 3 & 0.399558 & 0.313149 & 0.485641 &
0.326803 & 0.363762 & 0.305782 \\
\hline
\end{tabular}
\end{center}
%\vspace{0.3cm}
\begin{center}
{Table $19$. Quark mass critical exponent at the Banks-Zaks critical point for 
the $\MOMq$, $\MOMg$ and $\MOMh$ schemes at two and three loops for quarks in 
the $2S$ representation.}
\end{center}
\end{table}}

\clearpage 
{\begin{table}
\begin{center}
\begin{tabular}{|c|c||r|r|r||r|r|r|}
\hline
& $2A$ & & $\MSbars$ & & & $\mMOM$ & \\
\hline
$\Nc$ & $\Nf$ & $a_2$ & $a_3$ & $a_4$ & 
$a_2$ & $a_3$ & $a_4$ \\
\hline
4 & 6 & 0.172414 & 0.052865 & 0.061243 &
0.172414 & 0.044308 & 0.038398 \\
4 & 7 & 0.070796 & 0.034771 & 0.039931 &
0.070796 & 0.029895 & 0.028047 \\
4 & 8 & 0.035714 & 0.022840 & 0.025409 &
0.035714 & 0.020324 & 0.020083 \\
4 & 9 & 0.017937 & 0.013814 & 0.014662 &
0.017937 & 0.012777 & 0.012908 \\
4 & 10 & 0.007194 & 0.006401 & 0.006518 &
0.007194 & 0.006164 & 0.006212 \\
\hline
\end{tabular}
\end{center}
%\vspace{0.3cm}
\begin{center}
{Table $20$. Location of Banks-Zaks critical points for $\MSbar$ and $\mMOM$ at 
two, three and four loops for quarks in the $2A$ representation.}
\end{center}
\end{table}}

{\begin{table}
\begin{center}
\begin{tabular}{|c|c||r|r||r|r||r|r|}
\hline
& $2A$ & & $\MOMq$ & & $\MOMg$ & & $\MOMh$ \\
\hline
$\Nc$ & $\Nf$ & $a_2$ & $a_3$ & $a_2$ & 
$a_3$ & $a_2$ & $a_3$ \\
\hline
4 & 6 & 0.172414 & 0.036860 & 0.172414 & 0.032342 &
0.172414 & 0.045374 \\
4 & 7 & 0.070796 & 0.027053 & 0.070796 & 0.022389 &
0.070796 & 0.031004 \\
4 & 8 & 0.035714 & 0.019325 & 0.035714 & 0.015630 &
0.035714 & 0.021179 \\
4 & 9 & 0.017937 & 0.012543 & 0.017937 & 0.010258 &
0.017937 & 0.013285 \\
4 & 10 & 0.007194 & 0.006161 & 0.007194 & 0.005338 &
0.007194 & 0.006332 \\
\hline
\end{tabular}
\end{center}
%\vspace{0.3cm}
\begin{center}
{Table $21$. Location of Banks-Zaks critical points for $\MOMq$, $\MOMg$ and 
$\MOMh$ at two and three loops for quarks in the $2A$ representation.}
\end{center}
\end{table}}

{\begin{table}
\begin{center}
\begin{tabular}{|c|c||r|r|r||r|r|r|}
\hline
& $2A$ & & $\MSbars$ & & & $\mMOM$ & \\
\hline
$\Nc$ & $\Nf$ & $\omega_2$ & $\omega_3$ & $\omega_4$ & 
$\omega_2$ & $\omega_3$ & $\omega_4$ \\
\hline
4 & 6 & 2.298851 & 1.193609 & 1.109724 &
2.298851 & 1.029719 & 1.022181 \\
4 & 7 & 0.755162 & 0.559626 & 0.511494 &
0.755162 & 0.503114 & 0.508341 \\ 
4 & 8 & 0.285714 & 0.248588 & 0.229893 &
0.285714 & 0.232661 & 0.233704 \\
4 & 9 & 0.095665 & 0.090611 & 0.086504 &
0.095665 & 0.087749 & 0.087236 \\
4 & 10 & 0.019185 & 0.018951 & 0.018660 &
0.019185 & 0.018791 & 0.018680 \\
\hline
\end{tabular}
\end{center}
%\vspace{0.3cm}
\begin{center}
{Table $22$. Critical exponent $\omega$ for the Banks-Zaks critical point for 
$\MSbar$ and $\mMOM$ at two, three and four loops for quarks in the $2A$ 
representation.}
\end{center}
\end{table}}

{\begin{table}
\begin{center}
\begin{tabular}{|c|c||r|r||r|r||r|r|}
\hline
& $2A$ & & $\MOMq$ & & $\MOMg$ & & $\MOMh$ \\
\hline
$\Nc$ & $\Nf$ & $\omega_2$ & $\omega_3$ & $\omega_2$ & 
$\omega_3$ & $\omega_2$ & $\omega_3$ \\
\hline
4 & 6 & 2.298851 & 0.877863 & 2.298851 & 0.781571 &
2.298851 & 1.050754 \\
4 & 7 & 0.755162 & 0.466867 & 0.755162 & 0.402103 &
0.755162 & 0.516585 \\
4 & 8 & 0.285714 & 0.225543 & 0.285714 & 0.195358 &
0.285714 & 0.238387 \\
4 & 9 & 0.095665 & 0.087014 & 0.095665 & 0.078133 &
0.095665 & 0.089231 \\
4 & 10 & 0.019185 & 0.018789 & 0.019185 & 0.017908 &
0.019185 & 0.018909 \\
\hline
\end{tabular}
\end{center}
%\vspace{0.3cm}
\begin{center}
{Table $23$. Critical exponent $\omega$ for the Banks-Zaks critical point for 
the $\MOMq$, $\MOMg$ and $\MOMh$ schemes at two and three loops for quarks in 
the $2A$ representation.}
\end{center}
\end{table}}

\clearpage
{\begin{table}
\begin{center}
\begin{tabular}{|c|c||r|r|r||r|r|r|}
\hline
& $2A$ & & $\MSbars$ & & & $\mMOM$ & \\
\hline
$\Nc$ & $\Nf$ & $\rho_2$ & $\rho_3$ & $\rho_4$ & 
$\rho_2$ & $\rho_3$ & $\rho_4$ \\
\hline
4 & 6 & 9.782501 & 1.381815 & 0.292995 &
11.318371 & 1.566192 & 1.377240 \\
4 & 7 & 2.191767 & 0.695302 & 0.435137 &
2.484143 & 0.769888 & 0.703235 \\
4 & 8 & 0.801977 & 0.401949 & 0.368304 &
0.884885 & 0.429906 & 0.414671 \\
4 & 9 & 0.330860 & 0.228000 & 0.231646 &
0.353918 & 0.235533 & 0.235585 \\
4 & 10 & 0.116993 & 0.101120 & 0.102557 &
0.121047 & 0.101969 & 0.102620 \\
\hline
\end{tabular}
\end{center}
%\vspace{0.3cm}
\begin{center}
{Table $24$. Quark mass critical exponent at the Banks-Zaks critical point for 
the $\MSbar$ and $\mMOM$ schemes at two, three and four loops for quarks in the
$2A$ representation.}
\end{center}
\end{table}}

{\begin{table}[ht]
\begin{center}
\begin{tabular}{|c|c||r|r||r|r||r|r|}
\hline
& $2A$ & & $\MOMq$ & & $\MOMg$ & & $\MOMh$ \\
\hline
$\Nc$ & $\Nf$ & $\rho_2$ & $\rho_3$ & $\rho_2$ & 
$\rho_3$ & $\rho_2$ & $\rho_3$ \\
\hline
4 & 6 & 7.054427 & 0.805527 & 12.794194 & 1.154631 &
5.180018 & 0.582067 \\
4 & 7 & 1.882686 & 0.560447 & 3.197151 & 0.730752 &
1.566644 & 0.491650 \\
4 & 8 & 0.761721 & 0.375009 & 1.184460 & 0.452628 &
0.681294 & 0.353174 \\
4 & 9 & 0.330392 & 0.225207 & 0.459282 & 0.253314 &
0.310104 & 0.219754 \\
4 & 10 & 0.118476 & 0.101235 & 0.142790 & 0.106250 &
0.115212 & 0.100632 \\
\hline
\end{tabular}
\end{center}
%\vspace{0.3cm}
\begin{center}
{Table $25$. Quark mass critical exponent at the Banks-Zaks critical point for 
the $\MOMq$, $\MOMg$ and $\MOMh$ schemes at two and three loops for quarks in 
the $2A$ representation.}
\end{center}
\end{table}}


\begin{thebibliography}{99}
\bibitem{1} D.J. Gross \& F.J. Wilczek, Phys. Rev. Lett. {\bf 30}
(1973), 1343.
%%CITATION = PRLTA,30,1343;%%
\bibitem{2} H.D. Politzer, Phys. Rev. Lett. {\bf 30} (1973), 1346.
%%CITATION = PRLTA,30,1346;%%
\bibitem{3} W.E. Caswell, Phys. Rev. Lett. {\bf 33} (1974), 244.
%%CITATION = PRLTA,33,244;%%
\bibitem{4} D.R.T. Jones, Nucl. Phys. {\bf B75} (1974), 531.
%%CITATION = NUPHA,B75,531;%%
\bibitem{5} T. Banks \& A. Zaks, Nucl. Phys. {\bf B196} (1982), 189.
%%CITATION = NUPHA,B196,189;%%
\bibitem{6} E. Farhi \& L. Susskind, Phys. Rept. {\bf 74} (1981), 277.
%%CITATION = PRPLC,74,277;%%"
\bibitem{7} K. Yamawaki, M. Bando \& K.-I. Matumoto, Phys. Rev. Lett. {\bf 56}
(1986), 1335. 
%%CITATION = PRLTA,56,1335;%%
\bibitem{8} T. Appelquist, G.T. Fleming \& E.T. Neil, Phys. Rev. {\bf D79}
(2009), 076010.
%%CITATION = ARXIV:0901.3766;%%
\bibitem{9} Z. Fodor, K. Holland, J. Kuti, D. N\'{o}gr\'{a}di \& C.
Schr\"{o}der, Phys. Lett. {\bf B703} (2011), 348.
%%CITATION = ARXIV:1104.3124;%%
\bibitem{10} T. Karavirta, J. Rantaharju, K. Rummukainen \& K. Tuominen, 
JHEP {\bf 1205} (2012), 003
%%CITATION = ARXIV:1111.4104;%%
\bibitem{11} T. DeGrand, Y. Shamir \& B. Svetitsky, Phys. Rev. {\bf D88} 
(2013), 054505.
%%CITATION = ARXIV:1307.2425;%%
\bibitem{12} M. Hayakawa, K.-I. Ishikawa, S. Takeda \& N. Yamada,
Phys. Rev. {\bf D88} (2013), 094504.
%%CITATION = ARXIV:1307.6997;%%
\bibitem{13} M. Tomii, M. Hayakawa, K.-I. Ishikawa, S. Takeda \& N. Yamada,
PoS LATTICE2013 (2014), 068.
%%CITATION = ARXIV:1311.0099;%%
\bibitem{14} A. Cheng, A. Hasenfratz, Y. Liu, G. Petropoulos \& D. Schaich, 
Phys. Rev. {\bf D90} (2014), 014509.
%%CITATION = 1401.0195;%%
\bibitem{15} M.P. Lombardo, K. Miura, T.J. Nunes da Silva \& E. Pallante, JHEP
{\bf 1412} (2014), 183. 
%%CITATION = 1410.0298;%%
\bibitem{16} M. Hopfer, C.S. Fischer \& R. Alkofer, JHEP {\bf 1411} (2014),
035.
%%CITATION = 1405.7031;%%
\bibitem{17} T.A. Ryttov, Phys. Rev. {\bf D90} (2014), 056007.
%%CITATION = ARXIV:1408.5841;%%
\bibitem{18} E. Egorian \& O.V. Tarasov, Theor. Math. Phys. {\bf 41} (1979),
863.
%%CITATION = TMPHA,41,863;%%
\bibitem{19} O.V. Tarasov, A.A. Vladimirov \& A.Yu. Zharkov, Phys. Lett.
{\bf B93} (1980) 429.
%%CITATION = PHLTA,B93,429;%%
\bibitem{20} S.A. Larin \& J.A.M. Vermaseren, Phys. Lett. {\bf B303} (1993),
334.
%%CITATION = PHLTA,B303,334;%%
\bibitem{21} T. van Ritbergen, J.A.M. Vermaseren \& S.A. Larin, Phys. Lett.
{\bf B400} (1997), 379.
%%CITATION = HEP-PH 9701390;%%
\bibitem{22} M. Czakon, Nucl. Phys. {\bf B710} (2005), 485.
%%CITATION = HEP-PH 0411261;%%
\bibitem{23} O. Nachtmann \& W. Wetzel, Nucl. Phys. {\bf B187} (1981), 333.
%%CITATION = NUPHA,B187,333;%%
\bibitem{24} R. Tarrach, Nucl. Phys. {\bf B183} (1981), 384.
%%CITATION = NUPHA,B183,384;%%
\bibitem{25} O.V. Tarasov, JINR preprint P2-82-900.
%%CITATION = JINR-P2-82-900 ETC.;%%
\bibitem{26} K.G. Chetyrkin, Phys. Lett. {\bf B404} (1997), 161.
%%CITATION = HEP-PH 9703278;%%
\bibitem{27} J.A.M. Vermaseren, S.A. Larin \& T. van Ritbergen, Phys. Lett.
{\bf B405} (1997), 327.
%%CITATION = HEP-PH 9703284;%%
\bibitem{28} S.A. Caveny \& P.M. Stevenson, hep-ph/9705319. 
%%CITATION = HEP-PH 9705319;%%
\bibitem{29} T.A. Ryttov \& R. Shrock, Phys. Rev. {\bf D86} (2012), 065032.
%%CITATION = ARXIV:1206.2366;%%
\bibitem{30} T.A. Ryttov \& R. Shrock, Phys. Rev. {\bf D86} (2012), 085005.
%%CITATION = ARXIV:1206.6895;%%
\bibitem{31} R. Shrock, Phys. Rev. {\bf D89} (2014), 045019.
%%CITATION = ARXIV:1311.5268;%%
\bibitem{32} R. Shrock, Phys. Rev. {\bf D90} (2014), 045011.
%%CITATION = ARXIV:1405.6244;%%
\bibitem{33} G. Choi \& R. Shrock, Phys. Rev. {\bf D90} (2014), 125029.
%%CITATION = ARXIV:1411.6645;%%
\bibitem{34} T.A. Ryttov, Phys. Rev. {\bf D89} (2014), 016013.
%%CITATION = ARXIV:1309.3987;%%
\bibitem{35} T.A. Ryttov, Phys. Rev. {\bf D89} (2014), 056001.
%%CITATION = ARXIV:1311.0848;%%
\bibitem{36} G. Martinelli, C. Pittori, C.T. Sachrajda, M. Testa \& A.
Vladikas, Nucl. Phys. {\bf B445} (1995), 81.
%%CITATION = HEP-LAT 9411010;%%
\bibitem{37} E. Franco \& V. Lubicz, Nucl. Phys. {\bf B531} (1998), 641.
%%CITATION = HEP-LAT 9803491;%%
\bibitem{38} L. von Smekal, K. Maltman \& A. Sternbeck, Phys. Lett. {\bf B681}
(2009), 336.
%%CITATION = ARXIV:0903.1696;%%
\bibitem{39} K.G. Chetyrkin \& A. R\'{e}tey, Nucl. Phys. {\bf B583} (2000), 3.
%%CITATION = HEP-PH 9910332;%%
\bibitem{40} J.A. Gracey, Nucl. Phys. {\bf B662} (2003), 247.
%%CITATION = HEP-PH 0304113;%%
\bibitem{41} J.A. Gracey, J. Phys. {\bf A46} (2013), 225403.
%%CITATION = ARXIV:1304.5347;%%
\bibitem{42} J.C. Taylor, Nucl. Phys. {\bf B33} (1971), 436.
%%CITATION = NUPHA,B33,436;%%
\bibitem{43} W. Celmaster \& R.J. Gonsalves, Phys. Rev. Lett. {\bf 42} (1979),
1435.
%%CITATION = PRLTA,42,1435;%%
\bibitem{44} W. Celmaster \& R.J. Gonsalves, Phys. Rev. {\bf D20} (1979), 1420.
%%CITATION = PHRVA,D20,1420;%%
\bibitem{45} J.A. Gracey, Phys. Rev. {\bf D84} (2011), 085011.
%%CITATION = ARXIV:1108.4806;%%
\bibitem{46} S.A. Larin, Phys. Lett. {\bf B303} (1993), 113.
%%CITATION = HEP-PH 9302240;%%
\bibitem{47} K.G. Chetyrkin, A.L. Kataev \& F.V. Tkachov, Nucl. Phys. {\bf
B174} (1980), 345.
%%CITATION = NUPHA,B174,345;%%
\bibitem{48} A.A. Vladimirov, Theor. Math. Phys. {\bf 43} (1980), 417.
%%CITATION = TMPHA,43,417;%%
\bibitem{49} K.G. Chetyrkin, A.L. Kataev \& F.V. Tkachov, Nucl. Phys. {\bf
B174} (1980), 345.
%%CITATION = NUPHA,B174,345;%%
\bibitem{50} P.A. Baikov, K.G. Chetyrkin \& J.H. K\"{u}hn, JHEP {\bf 1410}
(2014), 76.
%%CITATION = ARXIV:1402.6611;%%
\bibitem{51} P.A. Baikov \& K.G. Chetyrkin, Nucl. Phys. {\bf B837} (2010), 186.
%%CITATION = ARXIV:1004.1153;%%
\bibitem{52} K.G. Chetyrkin \& T. Seidensticker, Phys. Lett. {\bf B495} (2000),
74.
%%CITATION = HEP-PH 0008094;%%
\bibitem{53} A.D. Kennedy, J. Math. Phys. {\bf 22} (1981), 1330.
%%CITATION = JMAPA,22,1330;%%
\bibitem{54} A. Bondi, G. Curci, G. Paffuti \& P. Rossi, Ann. Phys. {\bf 199}
(1990), 268.
%%CITATION = APNYA,199,268;%%
\bibitem{55} A.N. Vasil'ev, S.\'{E}. Derkachov \& N.A. Kivel, Theor. Math.
Phys. {\bf 103} (1995), 487.
%%CITATION = TMPHA,103,487;%%
\bibitem{56} S. Laporta, Int. J. Mod. Phys. {\bf A15} (2000), 5087.
%%CITATION = HEP-PH 0207004;%%
\bibitem{57} A.I. Davydychev, J. Phys. {\bf A25} (1992), 5587.
%%CITATION = JPAGB,A25,5587;%%
\bibitem{58} N.I. Usyukina \& A.I. Davydychev, Phys. Atom. Nucl. {\bf 56}
(1993), 1553.
%%CITATION = HEP-PH 9307327;%%
\bibitem{59} N.I. Usyukina \& A.I. Davydychev, Phys. Lett. {\bf B332} (1994),
159.
%%CITATION = HEP-PH 9402223;%%
\bibitem{60} T.G. Birthwright, E.W.N. Glover \& P. Marquard, JHEP {\bf 0409}
(2004), 042.
%%CITATION = HEP-PH 0407343;%%
\bibitem{61} L.G. Almeida \& C. Sturm, Phys. Rev. {\bf D82} (2010), 054017.
%%CITATION = 1004.4613;%%
\bibitem{62} C. Studerus, Comput. Phys. Commun. {\bf 181} (2010), 1293.
%%CITATION = 0912.2546;%%
\bibitem{63} J.A.M. Vermaseren, math-ph/0010025.
%%CITATION = MATH-PH 0010025;%%
\bibitem{64} M. Tentyukov \& J.A.M. Vermaseren, Comput. Phys. Commun. {\bf 181}
(2010), 1419.
%%CITATION = HEP-PH 0702279;%%
\bibitem{65} P. Nogueira, J. Comput. Phys. {\bf 105} (1993), 279.
%%CITATION = JCTPA,105,279;%%
\bibitem{66} O.V. Tarasov \& A.A. Vladimirov, Sov. J. Nucl. Phys. {\bf 25}
(1977), 585.
%%CITATION = SJNCA,25,585;%%
\bibitem{67} O.V. Tarasov \& D.V. Shirkov, Sov. J. Nucl. Phys. {\bf 51} (1990),
877.
%%CITATION = SJNCA,51,877;%%
\bibitem{68} G. 't Hooft, Nucl. Phys. {\bf B61} (1973), 455.
%%CITATION = NUPHA,B197,477;%%
\bibitem{69} J.M. Bell \& J.A. Gracey, Phys. Rev. {\bf D88} (2013), 085027.
%%CITATION = 1310.0243;%%
\bibitem{70} G. 't Hooft, Nucl. Phys. {\bf B190} (1981), 455.
%%CITATION = NUPHA,B190,455;%%
\bibitem{71} A.S. Kronfeld, G. Schierholz \& U.J. Wiese, Nucl. Phys. {\bf B293}
(1987), 461.
%%CITATION = NUPHA,B293,461;%%
\bibitem{72} A.S. Kronfeld, M.L. Laursen, G. Schierholz \& U.J. Wiese, Phys.
Lett. {\bf B198} (1987), 516.
%%CITATION = PHLTA,B198,516;%%
\bibitem{73} M. Gorbahn \& S. J\"{a}ger, Phys. Rev. {\bf D82} (2010), 114001.
%%CITATION = 1004.3997;%%
\bibitem{74} K.G. Wilson \& M.E. Fisher, Phys. Rev. Lett. {\bf 28} (1972), 240.
%%CITATION = PRLTA,28,240;%%
\bibitem{75} K.G. Wilson, Phys. Rev. {\bf B4} (1971), 3174.
%%CITATION = PHRVA,B4,3174;%%
\bibitem{76} K.G. Wilson, Phys. Rev. {\bf B4} (1971), 3184.
%%CITATION = PHRVA,B4,3184;%%
\bibitem{77} K.G. Wilson, Phys. Rev. Lett. {\bf 28} (1972), 548.
%%CITATION = PRLTA,28,548;%%
\bibitem{78} J.A. Gracey, paper in preparation. 
\bibitem{79} G. 't Hooft, Subnucl. Ser. {\bf 15} (1979), 943.
%%CITATION = SUSEE,15,943;%%
\end{thebibliography}
\end{document}